\documentclass{ws-rv9x6}
\usepackage{ws-rv-van}             
\makeindex
\def\minifig#1#2#3{\parbox{#1}{{#2}\figsubcap{#3}}}
\begin{document}

\chapter{The International Linear Collider}

\author[]{Marco Battaglia}

\address{Department of Physics, University of California at Berkeley and\\
Lawrence Berkeley National Laboratory\\
Berkeley, CA 94720, USA \\
MBattaglia@lbl.gov}

\begin{abstract}
The International Linear Collider (ILC) is the next large scale project in 
accelerator particle physics. Colliding electrons with positrons at energies
from 0.3~TeV up to about 1~TeV, the ILC is expected to provide the accuracy 
needed to complement the LHC data and extend the sensitivity to new phenomena 
at the high energy frontier and answer some of the fundamental questions in 
particle physics and in its relation to Cosmology. This paper reviews some 
highlights of the ILC physics program and of the major challenges for the 
accelerator and detector design.
\end{abstract}

\body

\section{Introduction}\label{sec0}

Accelerator particle physics is completing a successful cycle of precision
tests of the Standard Model of electro-weak interactions (SM). After the 
discovery of the $W$ and $Z$ bosons at the $Sp\bar{p}S$ hadron collider at 
CERN, the concurrent operation of hadron and $e^+e^-$ colliders has provided 
a large set of precision data and new observations. 
Two $e^+e^-$ colliders, the SLAC Linear Collider (SLC) at the Stanford Linear
Accelerator Center (SLAC) and the Large Electron Positron (LEP) collider at 
the European Organization for Nuclear Research (CERN), operated throughout the 
1990's and enabled the study of the properties of the $Z$ boson in great detail. 
Operation at LEP up to 209~GeV, the highest collision energy ever achieved in 
electron-positron collisions, provided detailed information on the properties of 
$W$ bosons and the strongest lower bounds on the mass of the Higgs boson and of 
several supersymmetric particles. The collision of point-like, elementary 
particles at a well-defined and tunable energy offers advantages for precision 
measurements, as those conducted at LEP and SLC, over proton colliders. 
On the other hand experiments at hadron machines, such as the Tevatron $p \bar p$ 
collider at Fermilab, have enjoyed higher constituent energies. The CDF and D0 
experiments eventually observed the direct production of top quarks, whose mass had 
been predicted on the basis of precision data obtained at LEP and SLC.

While we await the commissioning and operation of the LHC $pp$ collider at CERN, 
the next stage in experimentation at lepton colliders is actively under study.
For more than two decades, studies for a high-luminosity accelerator, able to 
collide electrons with positrons at energies of the order of 1~TeV, are being 
carried out world-wide. 

\section{The path towards the ILC}\label{sec1}

The concept of an $e^+e^-$ linear collider dates back to a paper by 
Maury Tigner~\cite{Tigner:1965} published in 1965, when the physics potential 
of $e^+e^-$ collisions had not yet been appreciated in full. This seminal paper 
envisaged collisions at 3-4~GeV with a luminosity competitive with that 
of the SPEAR ring at SLAC, i.e.\ $3 \times 10^{30}$~cm$^{-2}$~s$^{-1}$.
{\it A possible scheme to obtain $e^-e^-$ and $e^+e^-$ collisions at energies
of hundreds of GeV} is the title of a paper~\cite{Amaldi:1976} by Ugo Amaldi 
published a decade later in 1976, which sketches the linear collider concept 
with a design close to that now developed for the ILC. 
\begin{figure}
\centerline{\psfig{file=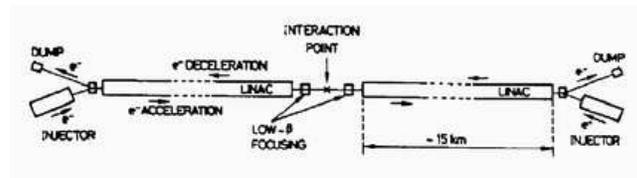,width=8.5cm}}
\caption{The linear collider layout as sketched in 1975 in one of the figures of 
Ref~\cite{Amaldi:1976}. The paper discussed the possibility to achieve  
$e^-e^-$ and $e^+e^-$ collisions at 0.3~TeV using superconducting linacs with a
gradient of 10~MV/m.}
\label{fig1}
\end{figure}
The parameters for a linear collider, clearly recognised as the successors of 
$e^+e^-$ storage rings on the way to high energies, were discussed by Burt Richter 
at the IEEE conference in San Francisco in 1979~\cite{Richter:1979cq} and soon after
came the proposal for the {\it Single Pass Collider Project} which would become 
SLC at SLAC.

From 1985, the CERN Long Range Planning Committee considered an $e^+e^-$ 
linear collider, based on the CLIC~\cite{Schnell:1986ig} design, able to deliver 
collisions at 2~TeV with $10^{33}$~cm$^{-2}$~s$^{-1}$ luminosity, {\it vis-a-vis} 
a hadron collider, with proton-proton collisions at 16~TeV and luminosity of 
$1.4 \times 10^{33}$~cm$^{-2}$~s$^{-1}$, as a candidate for the new CERN project 
after LEP. That review process eventually led to the decision to build 
the LHC, but it marked an important step to establish the potential of a high 
energy $e^+e^-$ collider. It is important to note that it was through the 
contributions of several theorists, including John Ellis, Michael Peskin, Gordon Kane
and others, that the requirements 
in terms of energy and luminosity for a linear collider became clearer in the 
mid 1980's~\cite{Ahn:1988vj}. 
The SLC project gave an important proof of principle for 
a high energy linear collider and the experience gained has shaped the subsequent 
designs in quite a significant way.

After a decade marked by important progress in the R\&D of the basic
components and the setup of advanced test facilities, designs of four different 
concepts emerged: TESLA, based on superconducting RF cavities, the NLC/JLC-X, 
based on high frequency (11.4~GHz) room-temperature copper cavities, JLC-C, 
based on lower frequency (5.7~GHz) conventional cavities and CLIC, a multi-TeV 
collider based on a different beam acceleration technique, the two-beam scheme 
with transfer structures operating at 30~GHz. Accelerator R\&D had reached
the maturity to assess the technical feasibility of a linear collider project 
and take an informed choice of the most advantageous RF technology. 
The designs were considered by the International Linear Collider Technical 
Review Committee (ILC-TRC), originally formed in 1994 and re-convened by the 
International Committee for Future Accelerators (ICFA) in 2001 under the 
chairmanship of Greg~A.~Loew. The ILC-TRC assessed their status using common 
criteria, identified outstanding items needing R\&D effort and suggested areas 
of collaboration. The TRC report was 
released in February 2003~\cite{trc} and the committee found that there were 
{\it no insurmountable show-stoppers to build TESLA, NLC/JLC-X or JLC-C in the 
next few years and CLIC in a more distant future, given enough resources}. 
Nonetheless, significant R\&D remained to be done. At this stage, it became clear that, 
to make further progress, the international effort towards a linear collider should be
focused on a single design. ICFA gave mandate to an 
International Technology Recommendation Panel (ITRP), chaired by Barry Barish, 
to make a definite recommendation for a RF technology that would be the basis 
of a global project. In August~2004 the ITRP made the 
recommendation in favour of superconducting RF cavities~\cite{itrp}. 
The technology choice, 
which was promptly accepted by all laboratories and groups involved in the R\&D 
process, is regarded as a major step towards the realization of the linear collider 
project. Soon after it, a truly world-wide, centrally managed design effort, 
the Global Design Effort (GDE)~\cite{gde}, a team of more than 60 persons, 
started, with the aim to produce an ILC Reference Design Report by beginning of 
2007 and an ILC Technical Design Report 
by end of 2008. The GDE responsibility now covers the detailed design 
concept, performance assessments, reliable international costing, 
industrialization plan, siting analysis, as well as detector concepts and 
scope. A further important step has been achieved with release of the Reference Design Report
in February 2007~\cite{rdr}. This report includes a preliminary value estimate of the cost 
for the ILC in its present design and at the present level of engineering and industrialisation. 
The value estimate is structured in three parts: 1.78 Billion ILC Value Units for site-related 
costs, such as those of tunneling in a specific region, 4.87 Billion ILC Value Units for the 
value of the high technology and conventional components and 13,000 person-years for the 
required supporting manpower. For this estimate the conversion factor is 
1 ILC Value Unit = 1~US Dollar = 0.83~Euro = 117~Yen. This estimate, which is comparable to 
the LHC cost, when the pre-existing facilities, such as the LEP tunnel, are included, provides 
guidance for optimisation of both the design and the R\&D to be done during the engineering 
phase, due to start in Fall 2007.

Technical progress was paralleled by increasing support for the ILC in the 
scientific community. At the 2001 APS workshop {\it The Future 
of Physics} held in Snowmass, CO, a consensus emerged for the ILC as the 
right project for the next large scale facility in particle physics. This 
consensus resonated and expanded in a number of statements by highly 
influential scientific advisory panels world-wide. The ILC role in the future 
of scientific research was recognised by the OECD Consultative Group on 
High Energy Physics~\cite{oecd}, while the DOE Office of Science ranked 
the ILC as its top mid-term project. More recently the EPP~2010 panel of the 
US National Academy of Sciences, in a report titled {\it Elementary Particle 
Physics in the 21$^{st}$ Century} has endorsed the ILC as the next major 
experimental facility to be built and its role in elucidating the physics at 
the high energy frontier, independently from the LHC findings~\cite{epp2010}.
Nowadays, the ILC is broadly regarded as the highest 
priority for a future large facility in particle physics, needed to extend 
and complement the LHC discoveries with the accuracy which is crucial to 
understand the nature of New Physics, test fundamental properties at the high 
energy scale and establish their relation to other fields in physical sciences, 
such as Cosmology. A matching program of 
physics studies and detector R\&D efforts has been in place for the past 
decade and it is now developing new, accurate and cost effective detector 
designs from proof of concepts towards that stage of engineering readiness, 
needed for being adopted in the ILC experiments.

\section{ILC Accelerator Parameters}\label{sec2}

\subsection{ILC Energy}\label{sec2.1}

The first question which emerges in defining the ILC parameters is 
the required centre-of-mass energy $\sqrt{s}$. It is here where 
we most need physics guidance to define the next thresholds at, 
and beyond, the electro-weak scale. The only threshold which, at 
present, is well defined numerically is that of top-quark pair 
production at $\sqrt{s} \simeq$ 350~GeV. Beyond it, there is a strong 
prejudice, supported by precision electro-weak and other data, 
that the Higgs boson should be light and new physics thresholds 
may exist between the electro-weak scale and approximately 1~TeV. 
If indeed the SM Higgs boson exists and the electro-weak data is not 
affected by new physics, its mass $M_H$ is expected to be below 
200~GeV as discussed in section~\ref{sec3.1}. 
Taking into account that the Higgs main production process 
is in association with a $Z^0$ boson, the maximum of the $e^+e^- \to
H^0 Z^0$ cross section varies from $\sqrt{s}$ = 240~GeV to 350~GeV 
for 120~GeV $< M_H <$ 200~GeV. On the other hand, we know that the 
current SM needs to be extended by some New Physics. Models of 
electroweak symmetry breaking contain new particles in the energy
domain below 1~TeV. More specifically, if Supersymmetry exists and 
it is responsible for the dark matter observed in the 
Universe, we expect that a significant fraction of the supersymmetric 
spectrum would be accessible at $\sqrt{s}$ = 0.5-1.0~TeV. 
In particular, the ILC should be able to study in detail those 
particles determining the dark matter relic density in the Universe
by operating at energies not exceeding 1~TeV, as discussed in 
section~\ref{sec3.2}.
Another useful perspective on the ILC energy is an analysis of the 
mass scale sensitivity for new physics vs. the $\sqrt{s}$ energy 
for lepton and hadron colliders in view of their synergy. The 
study of electro-weak processes at the highest available energy 
offers a window on mass scales well beyond its kinematic reach. 
A comparison of the mass-scale 
sensitivity for various new physics scenarios as 
a function of the centre-of-mass energy for $e^+e^-$ and $pp$ 
collisions is given in section~\ref{sec3.3}.
These and similar considerations, emerged in the course of the 
world-wide studies on physics at the ILC, motivate the choice of 
$\sqrt{s}$ = 0.5~TeV as the reference energy parameter, but requiring 
the ILC to be able to operate, with substantial luminosity, at 0.3~TeV 
as well and to be upgradable up to approximately 1~TeV. 

It is useful to consider these energies in an historical 
perspective. In 1954 Enrico Fermi gave a talk at the American Physical 
Society, of which he was chair, titled {\it What can we learn with 
high energy accelerators ?}. In that talk Fermi considered a proton 
accelerator with a radius equal to that of Earth and 2~T bending magnets, 
thus reaching a beam energy of $5 \times 10^{15}$~eV~\cite{Maiani:2001wi}.
Stanley Livingstone, who had built with Ernest O.~Lawrence the first 
circular accelerator at Berkeley in 1930, had formulated an empirical 
linear scaling law for the available centre-of-mass energy vs.\ the 
construction year and cost. Using Livingstone curve, Fermi predicted 
that such an accelerator could be built in 1994 at a cost of 
170~billion~\$. We have learned that, not only such accelerator could not 
be built, but accelerator physics has irrevocably fallen off the Livingstone 
curve, even in its revised version, which includes data up to the 1980's. 
As horizons expanded, each step has involved more and more technical 
challenges and has required more resources. The future promises to be 
along this same path. This underlines the need of coherent and responsible
long term planning while sustaining a rich R\&D program in both accelerator 
and detector techniques.

The accelerator envisaged by Enrico Fermi was a circular machine, 
as the almost totality of machines operating at the high energy frontier
still are. Now, as it is well known, charged particles undergoing a centripetal 
acceleration $a = v^2/R$ radiate at rate 
$P = \frac{1}{6 \pi \epsilon_0} \frac{e^2 a^2}{c^3} \gamma^4$.
If the radius $R$ is kept constant, the energy loss is the above rate $P$ 
times $t = 2 \pi R/v$, the time spent in the bending section of the accelerator. 
The energy loss for electrons is 
$W = 8.85 \times 10^{-5} \frac{E^4 \mathrm{(GeV^4)}}{R \mathrm{(km)}}$~MeV per turn 
while for protons is
$W = 7.8 \times 10^{-3} \frac{E^4 \mathrm{(TeV^4)}}{R \mathrm{(km)}}$~keV per turn.
Since the energy transferred per turn by the RF cavities to the beam is constant, 
$G \times 2 R \times F$, where $G$ is the cavity gradient and $F$ 
the tunnel fill factor, for each value of the accelerator ring 
radius $R$ there exists a maximum energy $E_{max}$ beyond which the energy loss 
exceeds the energy transferred. In practice, before this value of $E_{max}$ 
is reached, the real energy limit is set by the power dumped by the beam as 
synchrotron radiation. To make a quantitative 
example, in the case of the LEP ring, with a radius $R$ =4.3~km, a beam of 
energy $E_{beam}$=250~GeV, would lose 80~GeV/turn. Gunther Voss is thought 
to be the author of a plot comparing the guessed cost of a storage ring and a 
linear collider as a function of the $e^+e^-$ centre-of-mass energy. 
A $\sqrt{s}$=500~GeV storage ring, which would have costed an estimated 
14~billions~CHF in 1970's is aptly labelled as the 
{Crazytron}~\cite{Treille:2002iu}. LEP filled the last window 
of opportunity for a storage ring at the high energy frontier.
Beyond LEP-2 energies the design must be a linear collider, where no bending 
is applied to the accelerated particles. Still the accelerator length is 
limited by a number of constraints which include costs, alignment and siting. 
Therefore, technology still defines the maximum reachable energy at the ILC.

The ILC design is based on superconducting (s.c.\ ) radio-frequency (RF) 
cavities. While s.c.\ cavities had been considered already in the 1960's, 
it was Ugo Amaldi to first propose a fully s.c.\ linear collider in 
1975~\cite{Amaldi:1976}.
\begin{figure}
\centerline{\psfig{file=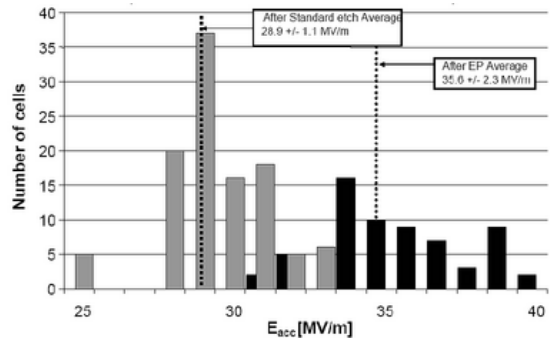,width=7.5cm}}
\caption{Distributions of gradients measured for pure niobium, nine-cell 
cavities. After electro-polishing an average gradient in excess to 35~MV/m 
has been obtained.}
\label{fig2}
\end{figure}
By the early 1990's, s.c.\ cavities equipped already one accelerator, 
TRISTAN at KEK in Japan, while two further projects were in progress,
CEBAF at Cornell and the LEP-2 upgrade at CERN. LEP-2 employed a total 
of 288~s.c.\ RF cavities, providing an average gradient of 7.2~MV/m. 
It was the visionary effort of Bjorn Wijk to promote, from 1990, 
the TESLA collaboration, with the aim to develop s.c.\ RF cavities 
pushing the gradient higher by a factor of five and the production costs 
down by a factor of four, thus reducing the cost per MV by a factor of 
twenty. Such reduction in cost was absolutely necessary to make a 
high energy collider, based on s.c.\ cavities, feasible. 
Within less than a decade 1.3~GHz, pure niobium cavities 
achieved gradients in excess to 35~MV/m. This opened the way to their
application to a $e^+e^-$ linear collider, able to reach centre-of-mass 
energies of the order of 1~TeV, as presented in detail in the TESLA 
proposal published in 2001~\cite{Brinkmann:2001qn} and recommended 
for the ILC by the ITRP in 2004~\cite{itrp}.
\begin{figure}
\centerline{\psfig{file=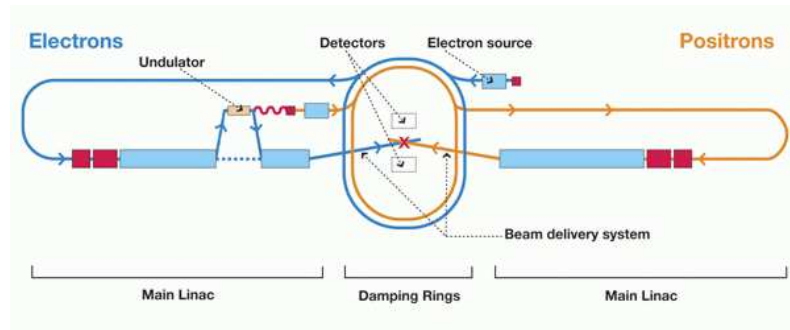,width=10.5cm}}
\caption{Schematic layout of the International Linear Collider. 
This diagram reflects the recommendations of the Baseline Configuration Document, 
a report published in December 2005 that outlines the general design of the machine. 
(Credit ILC Global Design Effort)}
\label{fig3}
\end{figure}

Today, the ILC baseline design aims at matching technical feasibility to
cost optimisation. One of the major goals of the current effort in the ILC 
design is to understand enough about its costs to provide a reliable indication 
of the scale of funding required to carry out the ILC project. Preparing a 
reliable cost estimate for a project to be carried out as a truly world-wide 
effort at the stage of a conceptual design that still lacks much of the detailed 
engineering designs as well as agreements for responsibility and cost sharing 
between the partners and a precise industrialisation plan is a great challenge. 
Still having good cost information as soon as possible, to initiate negotiations 
with the funding agencies is of great importance. 
An interesting example of the details entering in this process 
is the optimisation of the cost vs.\  cavity gradient for a 0.5~TeV collider. 
The site length scales inversely with the gradient $G$ while the cost of the 
cryogenics scales as $G^2/Q_0$ resulting in a minimum cost for a gradient of 
40~MW/m, corresponding to a tunnel length of 40~km, and a fractional cost 
increase of 10~\% for gradients of 25~MV/m or 57~MV/m. The chosen gradient 
of 35~MV/m, which is matched by the average performance of the most recent 
prototypes after electro-polishing, gives a total tunnel length of 
44~km with a cost increment from the minimum of just 1~\%. 

Beyond 1~TeV, the extension of conventional RF technology is more 
speculative. In order to attain collisions at energies in excess of about 
1~TeV, with high luminosity, significantly higher gradients are necessary.
As the gradient of s.c.\ cavities is limited below $\sim$~50~MV/m,  
other avenues should be explored. The CLIC technology~\cite{Assmann:2000hg}, 
currently being developed at CERN and elsewhere, may offer gradients of the order 
of 150~MV/m~\cite{wuensch}, allowing collision energies in the range 3-5~TeV 
with a luminosity of $10^{35}$~cm$^{-2}$~s$^{-1}$, which would support a 
compelling physics program~\cite{Battaglia:2004mw}.
\begin{figure}
\centerline{\psfig{file=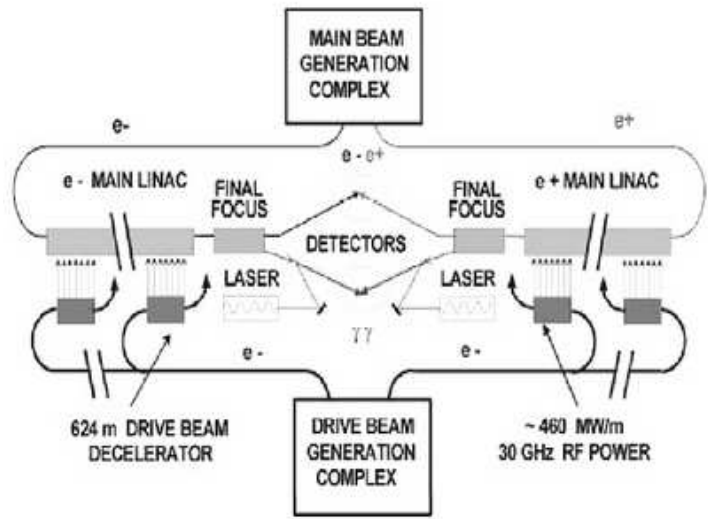,width=9.5cm}}
\caption{Schematics of the overall layout of the CLIC complex for $e^+e^-$ 
collisions at $\sqrt{s}$ = 3~TeV. (from Ref\cite{Battaglia:2004mw})}
\label{fig4}
\end{figure}
While RF cavities are limited to accelerating fields of order of 100-200~MV/m, 
or below, laser-wakefield accelerators are capable, in principle, of producing 
fields of 10-100~GV/m. Recently a 1~GeV $e^-$ beam has been accelerated over 
just 3.3~cm using a 40~TW peak-power laser pulse~\cite{Leemans:2006}, thus 
opening a possible path towards ultra-high energies in $e^+e^-$ collisions
in some more distant future.

\subsection{ILC Luminosity}\label{sec2.2}

The choice of a linear collider, rather than a circular storage ring, while 
solving the problem of the maximum reachable energy, introduces the challenge 
of achieving collisions with the required luminosity. The luminosity, 
$\cal{L}$, defined as the proportionality factor between the number of 
events produced and the process cross section $\sigma$, has requirements 
which depend on the typical values of s-channel cross sections and so scale 
as $1/s$. First luminosity requirements were already outlined in the 
1980s~\cite{Richter:1981uu,Amaldi:1987xt} 
as ${\cal{L}} \simeq \frac{2 E_{beam}}{\mathrm{TeV}} \times 
10^{33}$~cm$^{-2}$~s$^{-1}$, based on the estimated discovery potential.
But in the present vision of the ILC role in probing the high energy frontier
new requirements must be considered. One example is the precision study of 
electro-weak processes to look for deviations from the SM predictions, due to 
effect of new physics at high scales. The $e^+e^- \to b \bar b$ cross section 
at 1~TeV is just 96~fb, so this would corresponds to less than $10^3$ events 
per year at $10^{33}$~cm$^{-2}$~s$^{-1}$, which is certainly insufficient for 
the kind of precision measurements which we expect from the ILC. 
Another example is offered by one of the reactions most unique to the ILC: 
the double-Higgs production $e^+e^- \to HHZ$ sensitive to the Higgs self-coupling, 
which has a cross section of order of only 0.2~fb at 0.5~TeV. Therefore a 
luminosity of $10^{34}$~cm$^{-2}$~s$^{-1}$ or more is required as baseline 
parameter.
 
The luminosity can be expressed as a function of the accelerator parameters 
as:
\begin{equation}
 {\cal{L}} = f_{rep} n_b \frac{N^2}{4 \pi \sigma_x \sigma_y}.
\end{equation}
Now, since in a linear machine the beams are collided only once and then 
dumped, the collision frequency, $f_{rep}$, is small and high luminosity 
should be achieved by 
increasing the number of particles in a bunch $N$, the number of bunches $n_b$
and decreasing the transverse beam size $\sigma$. Viable values for $N$ are 
limited by wake-field 
effects and the ILC parameters have the same number of electrons in a bunch as 
LEP had, though it aims at a luminosity three orders of magnitude higher. 
Therefore, the increase must come from a larger number of bunches 
and a smaller transverse beam size. The generation of beams of small transverse 
size, their preservation during acceleration and their focusing to spots of 
nanometer size at the interaction region presents powerful challenges which the 
ILC design must solve. 
A small beam size also induces beam-beam interactions. On one hand the 
beam self-focusing, due to the electrostatic attraction of particles of 
opposite charges enhances the luminosity. But beam-beam interactions also 
result in an increase of beamstrahlung with a larger energy spread of the 
colliding particles, a degraded luminosity spectrum and higher backgrounds. 
Beamstrahlung is energy loss due to particle radiation triggered by the 
trajectory bending in the interactions with the charged particles in the 
incoming bunch~\cite{Noble:1986yz}. The mean beamstrahlung energy loss, 
which has to be minimised, is given by:
\begin{equation}
\delta_{BS} \simeq 0.86 \frac{e r_e^3}{2 m_0 c^2}
\frac{E_{cm}}{\sigma_z} \frac{N_b^2}{(\sigma_x+\sigma_y)^2}.
\end{equation}
Since the luminosity scales as $\frac{1}{\sigma_x \sigma_y}$, while the beamstrahlung 
energy loss scales as $\frac{1}{\sigma_x + \sigma_y}$, it is advantageous to choose 
a large beam aspect ratio, with the vertical beam size much smaller than the horizontal 
component. The parameter optimisation for luminosity can be further understood
by expressing the luminosity in terms of beam power 
$P = f_{rep} N E_{cm}$ = $\eta P_{AC}$ and beamstrahlung energy loss as:
\begin{equation}
{\cal{L}} \propto \frac{\eta P_{AC}}{E_{cm}} \sqrt{\frac{\delta_{BS}}{\epsilon_y}}H_D
\end{equation}
which highlights the dependence on the cavity efficiency $\eta$ and the total
power $P_{AC}$. The $H_D$ term is the pinch enhancement factor, that accounts for 
the bunch attraction in the collisions of oppositely charged beams. 
In summary, since the amount of available power is necessarily 
limited, the main handles on luminosity are $\eta$ and $\epsilon_y$. The 
efficiency for transferring power from the plug to the beam is naturally higher for 
s.c.\ than for conventional copper cavities, so more relaxed collision parameters can 
be adopted for a s.c.\ linear collider delivering the same luminosity. The main beam 
parameters for the ILC baseline design are given in Table~\ref{tab:params}.

\begin{table}[h]
\tbl{ILC baseline design beam parameters}
{\begin{tabular}{@{}ccc@{}} \toprule
Parameter & $\sqrt{s}$ & $\sqrt{s}$ \\
          & 0.5~TeV & 1.0~TeV \\ \colrule
Luminosity $L$ (10$^{34}$ cm$^{-2}$s$^{-1}$) & 2.0 & 2.8 \\
Frequency (Hz) & 5.0 & 5.0 \\
Nb.\ of particles (10$^{10}$) & 2.0 & 2.0 \\
Nb.\ of bunches $N_b$ & 2820 & 2820 \\
Bunch spacing (ns) & 308 & 308 \\
Vertical beam size $\sigma_y$ (nm) & 5.7 & 3.5 \\
Beamstrahlung Parameter $\delta_{BS}$ & 0.022 & 0.050 \\
$H_D$ & 1.7 & 1.5 \\ \botrule
\end{tabular}}
\label{tab:params}
\end{table}
 
\section{ILC Physics Highlights}\label{sec3}

The ILC physics program, as we can anticipate it at present, is broad and 
diverse, compelling and challenging. The ILC is being designed for operation 
at 0.5~TeV with the potential to span the largest range of collision energies, 
from the $Z^0$ peak at 0.091~TeV up to 1~TeV, collide electrons with positrons, 
but optionally also electrons with electrons, photons with photons and photons 
with electrons, and combine various polarization states of the electron and 
positron beams. Various 
reports discussing the linear collider physics case, including results of 
detailed physics studies, have been published in the last few 
years~\cite{Ahn:1988vj,Murayama:1996ec,Aguilar-Saavedra:2001rg,Abe:2001wn,Abe:2001gc,Dawson:2004xz,Battaglia:2004mw}.
Here, I shall focus on three of the main ILC physics themes: the 
detailed study of the Higgs boson profile, the determination of neutralino 
dark matter density in the Universe from accelerator data, and the sensitivity 
to new phenomena beyond the ILC kinematic reach, through the analysis of 
two-fermion production, at the highest $\sqrt{s}$ energy. 
Results discussed in the following have been obtained mostly using 
realistic, yet parametric simulation of the detector response.
Only few analyses have been carried out which include the full set of 
physics and machine-induced backgrounds on fully simulated and reconstructed 
events. With the progress of the activities of detector concepts and the 
definition of well-defined benchmark processes, this is becoming one of 
the priorities for the continuation of physics and detector studies.

\subsection{The Higgs Profile at the ILC}\label{sec3.1}

Explaining the origin of mass is one of the great scientific quests of our 
time. The SM addresses this question by the Higgs mechanism~\cite{Higgs}. 
The first direct manifestation of the Higgs mechanism through the Higgs
sector will be the existence of at least one Higgs boson. The observation
of a new spin-0 particle would represent a first sign that the Higgs 
mechanism of mass generation is indeed realised in Nature. This has 
motivated a large experimental effort, from LEP-2 
to the Tevatron and, soon, the LHC, actively backed-up by new and more 
accurate theoretical predictions. After a Higgs discovery, which we anticipate 
will be possible at the LHC, full validation of the Higgs mechanism can only 
be established by an accurate study of the Higgs boson production and decay 
properties. It is here where the ILC potential in precision physics will be 
crucial for the validation of the Higgs mechanism, through a detailed study 
of the Higgs profile~\cite{Heinemeyer:2005gs}.

The details of this study depend on the Higgs boson mass, $M_H$. In the SM, 
$M_H = \sqrt{2 \lambda} v$ where the Higgs field expectation 
value $v$ is determined as $(\sqrt{2}G_F)^{-1/2} \approx 246$~GeV, while 
the Higgs self-coupling $\lambda$ is not specified, leaving the mass as a 
free parameter. However, we have strong indications that $M_H$ must be 
light. The Higgs self-coupling behaviour at high energies~\cite{triv}, the 
Higgs field contribution to precision electro-weak data~\cite{ewwg:2005di} 
and the results of direct searches at LEP-2~\cite{Barate:2003sz} at 
$\sqrt{s} \ge$ 206~GeV, all point towards a light Higgs boson.
In particular, the study of precision electro-weak data, which are sensitive to 
the Higgs mass logarithmic contribution to radiative corrections, is based on 
several independent observables, including masses ($m_{top}$, $M_W$, $M_Z$), 
lepton and quark asymmetries at the $Z^0$ pole, $Z^0$ lineshape and partial 
decay widths. The fit to eighteen observables results in a 95\%~C.L. upper 
limit for the Higgs mass of 166~GeV, which becomes 199~GeV when the lower limit 
from the direct searches at LEP-2, $M_H >$ 114.4~GeV, is included. As a result, 
current data indicates that the Higgs boson mass should be in the range 
114~GeV $< M_H <$ 199~GeV. It is encouraging to observe that if the same fit 
is repeated, but excluding this time $m_{top}$ or $M_W$, the results for their 
values, 178$^{+12}_{-9}$~GeV and 80.361$\pm$0.020~GeV respectively, are in 
very good agreement with the those obtained the direct determinations, 
$m_{top}$ = 171.4$\pm$2.1~GeV and $M_W$ = 80.392$\pm$0.029~GeV.

At the ILC the Higgs boson can be observed in the Higgs-strahlung production 
process $e^+e^- \rightarrow HZ$ with $Z \rightarrow \ell^+\ell^-$, 
independent of its decay mode, by the distinctive peak in the di-lepton
recoil mass distribution. A data set of 500~fb$^{-1}$ at $\sqrt{s}$ = 350~GeV,
corresponding to four years of ILC running, 
provides a sample of 3500-2200 Higgs particles produced in the di-lepton $HZ$ 
channel, for $M_H$ = 120-200~GeV. Taking into account the SM backgrounds, 
dominated by $e^+e^- \rightarrow Z^0Z^0$ and $W^+W^-$ production, 
the Higgs boson observability is guaranteed up to its production kinematical 
limit, independent of its decays. This sets the ILC aside from the LHC, since 
the ILC sensitivity to the Higgs boson does not depend on its detailed properties.

\begin{figure}[h]%
\begin{center}
  \parbox{2.1in}{\epsfig{figure=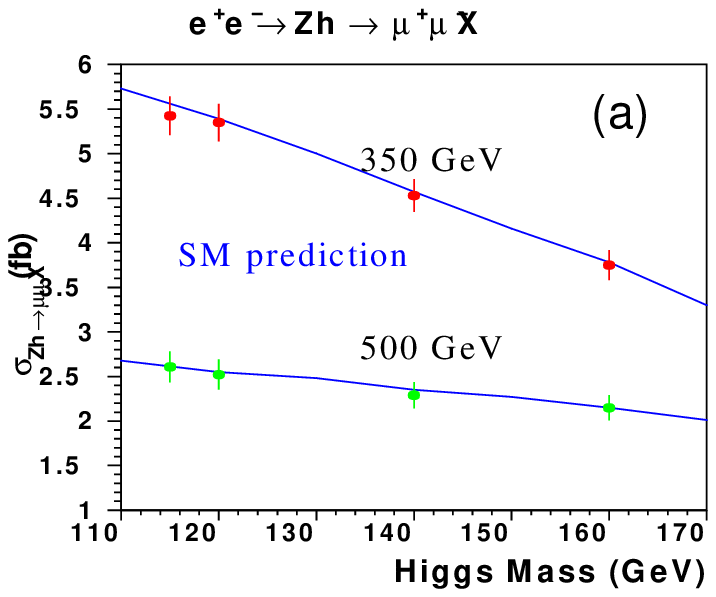,width=5.75cm,height=5.5cm}
  \figsubcap{a}}
  \hspace*{1pt}
  \parbox{2.1in}{\epsfig{figure=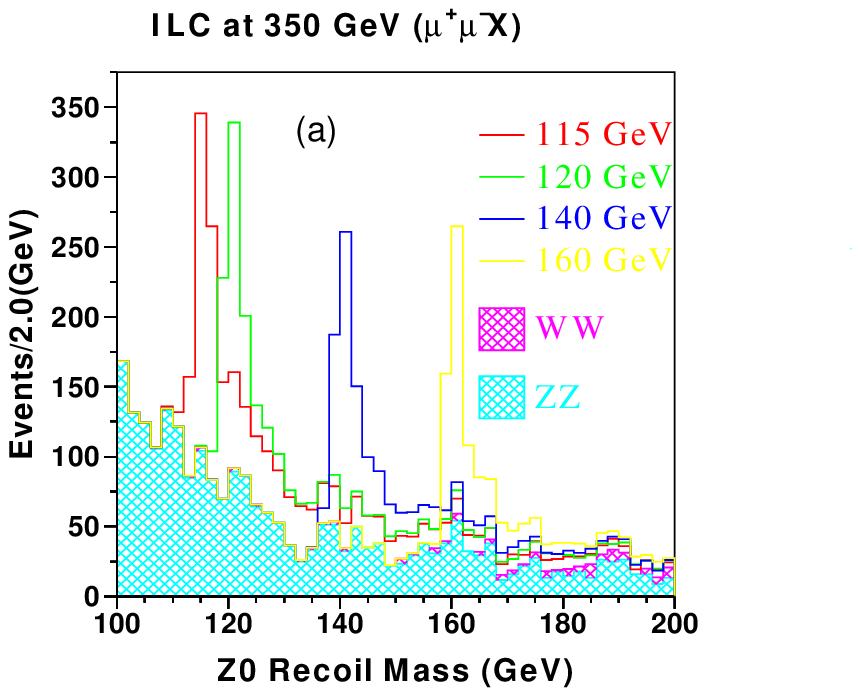,width=5.75cm,height=5.5cm}
  \figsubcap{b}}
  \caption{The Higgs-strahlung process at the ILC.
  (a) $e^+e^- \to HZ$ cross section vs.\ $M_H$ for $\sqrt{s}$ = 0.35~TeV and 0.5~TeV, 
  (b) reconstructed $\mu^+ \mu^-$ recoil mass for various values of the Higgs boson 
mass (Credit: ALCPG Study Group).}%
  \label{fig5}%
\end{center}
\end{figure}
After observation of a new particle with properties compatible with those
of the Higgs boson, a significant experimental and theoretical effort will be
needed to verify that this is indeed the boson of the scalar 
field responsible for the electro-weak symmetry breaking and the generation 
of mass. Outlining the Higgs boson profile, through the determination of its 
mass, width, quantum numbers, couplings to gauge bosons and fermions and the
reconstruction of the Higgs potential, stands as a most challenging, yet 
compelling, physics program. The ILC, with its large data sets at different 
centre-of-mass energies and beam polarisation conditions, the high resolution
detectors providing unprecedented accuracy on the reconstruction of the event 
properties and the use of advanced analysis techniques, developed from those 
successfully adopted at LEP and SLC, promises to promote Higgs physics into 
the domain of precision measurements.
Since the Higgs mass $M_H$ is not predicted by theory, it is of great interest 
to measure it precisely. Once this mass, and thus $\lambda$, is fixed, the 
profile of the Higgs particle is uniquely determined in the SM. In most scenarios 
we expect the LHC to determine the Higgs mass with a good accuracy. At the ILC, 
this measurement can be refined by exploiting the kinematical characteristics 
of the Higgs-strahlung production process 
$e^+e^- \rightarrow Z^* \rightarrow H^0 Z^0$ where the $Z^0$ can be 
reconstructed in both its leptonic and hadronic decay modes.
The $\ell^+\ell^-$ recoil mass for leptonic $Z^0$ decays yields an accuracy of
110~MeV for 500~fb$^{-1}$ of data, without any requirement on the nature
of the Higgs decays. Further improvement can be obtained by explicitly 
selecting $H \rightarrow b \bar b$ ($WW$) for $M_H \le$($>$) 140~GeV.
Here a kinematical 5-C fit, imposing energy and momentum conservation and the 
mass of a jet pair to correspond to $M_Z$, achieves an accuracy of 40 to 
90~MeV for 120$< M_H <$ 180~GeV~\cite{hmass1}.

The total decay width of the Higgs boson is predicted to be too narrow
to be resolved experimentally for Higgs boson masses below the $ZZ$ threshold.
On the contrary, above $\simeq$~200 GeV, the total width can be measured 
directly from the reconstructed width of the recoil mass peak, as discussed 
below. For the lower mass range, indirect methods must be applied.
In general, the total width is given by 
$\Gamma_{tot}=\Gamma_X/\mathrm{BR}(H\to X)$. Whenever $\Gamma_X$ can be 
determined independently of the corresponding branching fraction, a measurement 
of $\Gamma_{tot}$ can be carried out. The most convenient choice is the extraction 
of $\Gamma_{H}$ from the measurements of the $WW$ fusion cross section and 
the $H\to WW^*$ decay branching fraction . A relative precision of 6\% to 13\% 
on the width of the Higgs boson can be obtained at the ILC with 
this technique, for masses between 120~GeV and 160~GeV.
\begin{figure}[h]%
\begin{center}
\centerline{\epsfig{figure=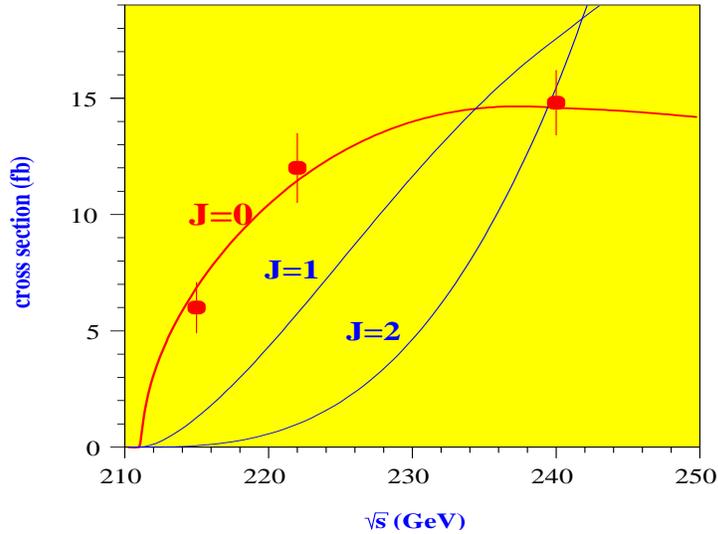,width=9.5cm,height=7.0cm}}
\caption{Determination of the Higgs boson spin from a scan of the 
$e^+e^- \to HZ$ cross section at threshold at the ILC 
(from Ref~\cite{Aguilar-Saavedra:2001rg}).}
\label{fig6}%
\end{center}
\end{figure}
The spin, parity and charge-conjugation quantum numbers $J^{PC}$ of Higgs
bosons can be determined at the ILC in a model-independent way. Already the 
observation of either $\gamma \gamma \rightarrow H$ production or 
$H \rightarrow \gamma\gamma$ decay sets $J \ne 1$ and $C=+$.
The angular dependence $\frac{d \sigma_{ZH}}{d \theta} \propto \sin^2 \theta$ 
and the rise of the Higgs-strahlung cross section:
\begin{equation}
\sigma_{ZH} \propto~\beta \sim~\sqrt{s-(M_H+M_Z)^2}
\end{equation}
allows to determine $J^P = 0^+$ and distinguish the SM Higgs 
from a $CP$-odd $0^{-+}$ state $A^0$, or a $CP$-violating mixture of the 
two~\cite{Miller:2001bi,Schumacher:2001ax}.
But where the ILC has a most unique potential is in verifying that the Higgs 
boson does its job of providing gauge bosons, quarks and leptons with their masses. 
This requires to precisely test the relation $g_{HXX} \propto {m_X}$ between 
the Yukawa couplings, $g_{HXX}$, and the corresponding particle masses, $m_X$.
\begin{figure}
\centerline{\epsfig{file=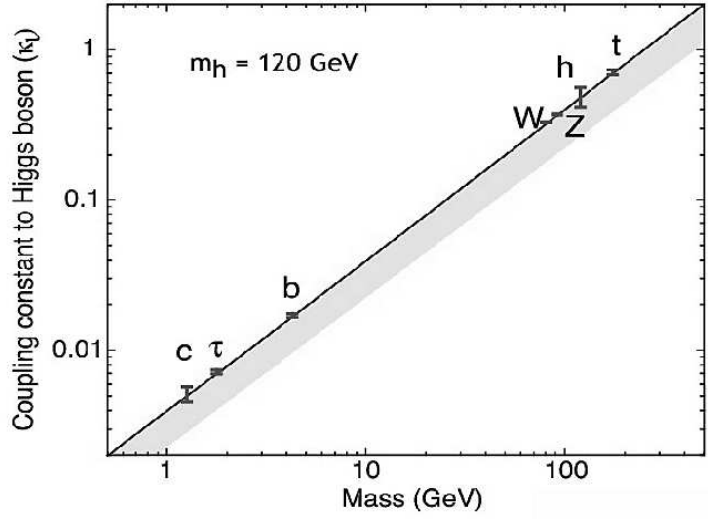,width=9.5cm,clip}}
\caption{Particle couplings to the Higgs field, for a 120~GeV boson, as a 
function of the particle masses. The error bars show the expected ILC accuracy 
in determining each of the couplings. The dark line is the SM prediction, while 
the shaded gray area shows the range of predictions from theories of new physics 
beyond the SM with extra dimensions (Credit: ACFA ILC Study Group).}
\label{fig7}
\end{figure}
In fact, the SM Higgs couplings to fermion pairs $g_{Hff} = m_f/v$ are fully 
determined by the fermion mass $m_f$. The corresponding decay partial widths 
only depend on these couplings and on the Higgs boson mass, QCD corrections 
do not represent a significant source of uncertainty~\cite{Djouadi:1995gt}.
Therefore, their 
accurate determination will represent a comprehensive test of the Higgs 
mechanism of mass generation~\cite{Hildreth:1993dx}.
Further, observing deviations of the measured 
values from the SM predictions will probe the structure of the Higgs sector 
and may reveal a non-minimal implementation of the Higgs model or the effect 
of new physics inducing a shift of the Higgs 
couplings~\cite{Carena:2001bg,Desch:2004cu,Battaglia:2004js}.
The accuracy of these measurements relies on the performances of jet flavour tagging 
and thus mostly on the Vertex Tracker, making this analysis an important benchmark 
for optimising the detector design. It is important to ensure that the ILC 
sensitivity extends over a wide range of Higgs boson masses and that a 
significant accuracy is achieved for most particle species. Here, the ILC adds 
the precision which establishes the key elements of the Higgs mechanism. It is 
important to point out that these tests are becoming more stringent now that 
the $B$-factories have greatly improved the determination of the $b$- and 
$c$-quark masses. When one of these studies was first presented in 
1999~\cite{Battaglia:1999re}, the $b$ quark mass was known to $\pm 0.11$~GeV
and the charm mass to $\pm 0.13$~GeV, with the expectation that $e^+e^-$ 
$B$-factory and LHC data could reduce these uncertainties by a factor of two 
by the time the ILC data would be analysed. Today, the analysis of a fraction 
of the BaBar data~\cite{Aubert:2004aw} has already brought these uncertainties 
down to 0.07~GeV for $m_b$ and, more importantly, 0.09~GeV for $m_c$, using 
the spectral moments technique in semi-leptonic $B$ decays, which had been 
pioneered on CLEO~\cite{Bauer:2002sh} and DELPHI data~\cite{Battaglia:2002tm}. 
Extrapolating to the anticipated total statistics to be collected at PEP-II 
and KEKB, we can now confidently expect that the $b$ quark mass should be known to 
better than $\pm 0.05$~GeV and the charm mass to better than $\pm 0.06$~GeV. This
translates into less than $\pm 0.4$~\% and $\pm 6.5$~\% relative uncertainty in 
computing the Higgs SM couplings to $b$ and $c$ quarks, respectively, and motivates 
enhanced experimental precision in the determination of these couplings at the ILC.
Detailed simulation shows that these accuracy can be matched by the 
ILC~\cite{Kuhl:2004ri,Barklow:2003hz}.

While much of the emphasis on the ILC capabilities in the study of the Higgs 
profile is for a light Higgs scenario, preferred by the current electro-weak 
data and richer in decay modes, the ILC has also the potential of precisely 
mapping out the Higgs boson properties for heavier masses. If the Higgs boson 
turns out to weigh of order 200~GeV, the 95\%~C.L.\ upper limit indicated by 
electro-weak fits, or even heavier, the analysis of the recoil mass in 
$e^+e^- \to HZ$ at $\sqrt{s}$ = 0.5~TeV allows to precisely determine $M_H$, 
$\Gamma_H$ and the Higgs-strahlung cross section. Even for $M_H$ = 240~GeV, the 
mass can be determined to a 10$^{-3}$ accuracy and, more importantly, the 
total width measured about 10\% accuracy. Decays of Higgs bosons 
produced in $e^+e^- \to H \nu \bar \nu$ give access to the Higgs couplings.
\begin{figure}[h]%
\begin{center}
  \parbox{2.1in}{\epsfig{figure=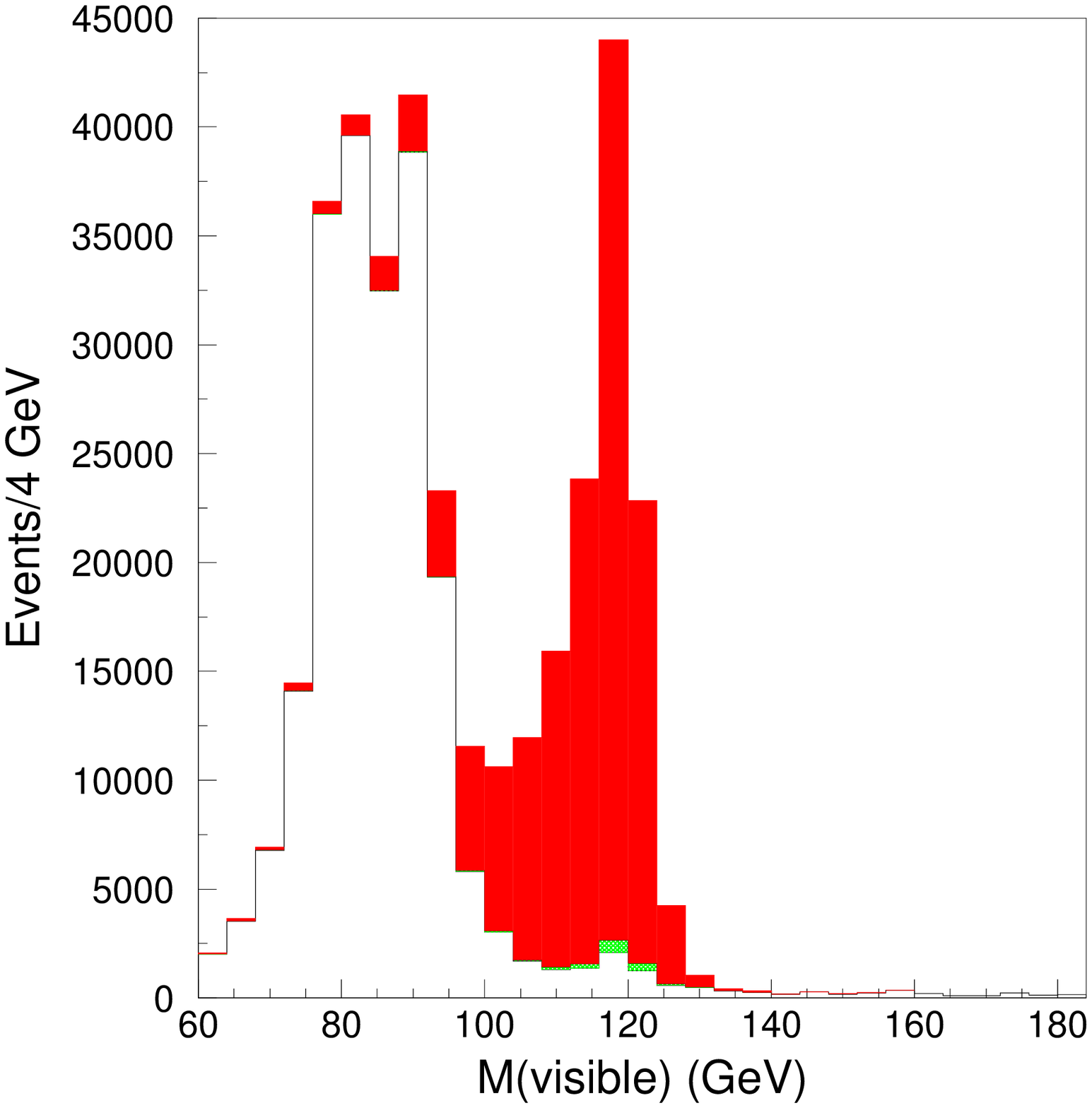,width=5.5cm,height=4.5cm}
  \figsubcap{a}}
  \hspace*{4pt}
  \parbox{2.1in}{\epsfig{figure=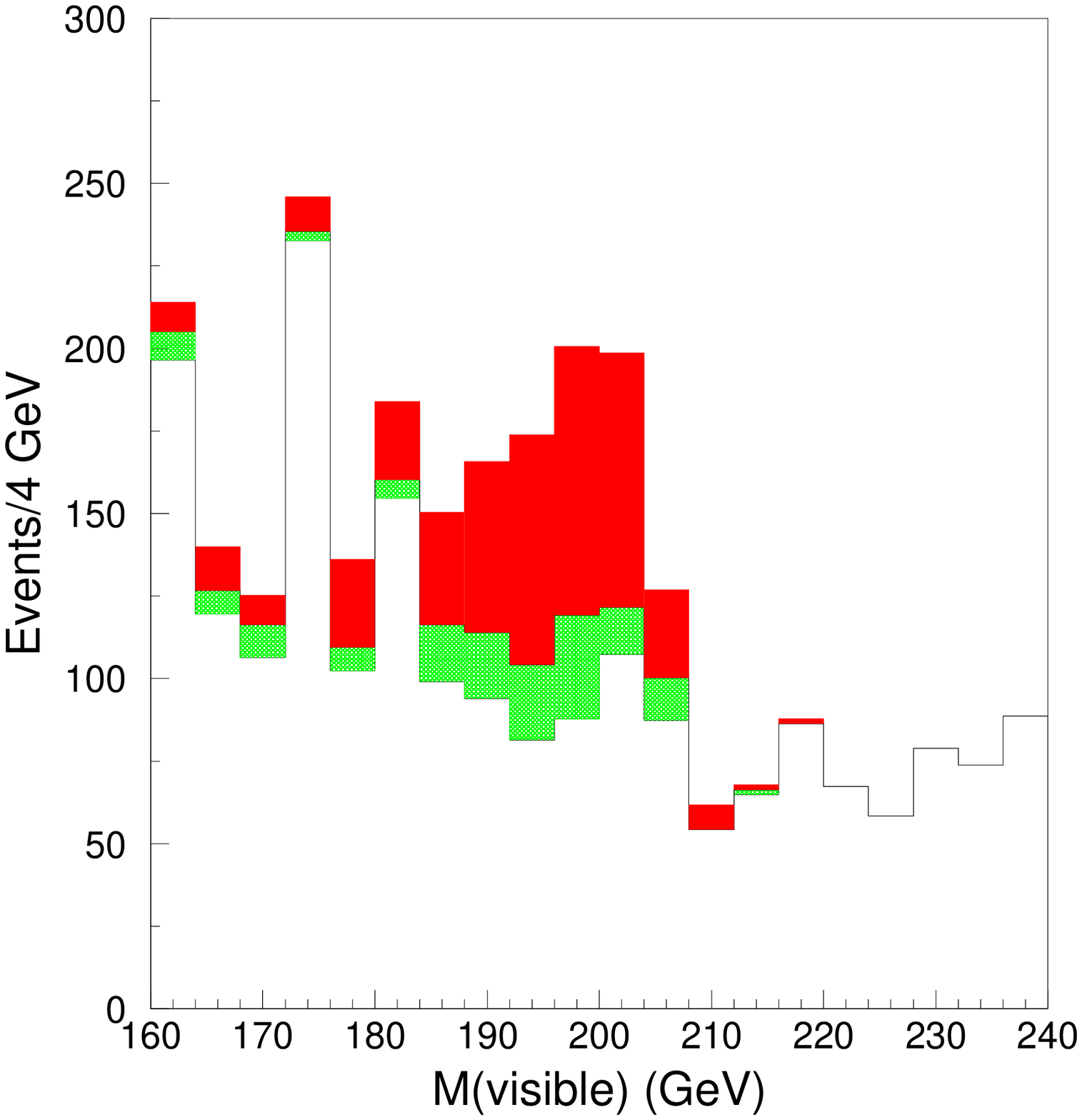,width=5.5cm,height=4.5cm}
  \figsubcap{b}}
  \caption{$H \to b \bar b$ signal after full event selection at the ILC for
  (a) $M_H$ = 120~GeV and
  (b) $M_H$ = 200~GeV (from Ref\cite{Barklow:2003hz}).}%
  \label{fig8}%
\end{center}
\end{figure}
The importance of the $WW$-fusion process $e^+e^- \to H^0 \nu \bar \nu$ to probe 
rare Higgs decays at higher energies, emerged in the physics study for a multi-TeV 
linear collider~\cite{Battaglia:2002gq}. Since this cross section increases as 
$log \frac{s}{M_H^2}$, it becomes dominant around $\sqrt{s}$ = 1~TeV. Detailed 
studies have been performed and show that 1~ab$^{-1}$ of data at $\sqrt{s}$ = 1~TeV,
corresponding to three to four years of ILC running,  
can significantly improve the determination of the Higgs couplings, especially for 
the larger values of $M_H$~\cite{Battaglia:2002av,Barklow:2003hz}.
$WW$ and $ZZ$ couplings can be determined with relative accuracies of 3~\% and 5~\% 
respectively, while the coupling to $b \bar b$ pairs, a rare decay with a branching
fraction of just $2 \times 10^{-3}$ at such large masses, can be determined to 
4~\% to 14~\% for 180~GeV $< M_H <$ 220~GeV. This measurement is of great importance,
since it would offer the only opportunity to learn about the fermion couplings of such 
an heavy Higgs boson, and it is unique to a linear collider. 

A most distinctive feature of the Higgs mechanism is the shape of the Higgs potential:
\begin{equation} 
V(\Phi) = - \frac{\mu^2}{2} \Phi^2 + \frac{\lambda}{4} \Phi^4
\end{equation}
with $v = \sqrt{\frac{\mu^2}{\lambda}}$. In the SM, the triple Higgs 
coupling, $g_{HHH} = 3 \lambda v$, is related to the Higgs mass, $M_H$, through 
the relation 
\begin{equation}
g_{HHH} = \frac{3}{2} \frac{M_H^2}{v}.
\end{equation} 
By determining $g_{HHH}$, the above relation 
can be tested. The ILC has access to the triple Higgs coupling through the 
double Higgs production processes $e^+e^- \to H H Z$ and 
$e^+e^- \to H H \nu\nu$~\cite{Djouadi:1999gv}. Deviations from the SM 
relation for the strength of the Higgs self-coupling arise in models with an extended 
Higgs sector~\cite{Kanemura:2004mg}. The extraction of $g_{HHH}$ is made 
difficult by their tiny cross sections and by the dilution effect, due to diagrams 
leading to the same double Higgs final states, but not sensitive to the triple Higgs 
vertex. This makes the determination of $g_{HHH}$ a genuine experimental 
{\it tour de force}.
\begin{figure}[h]%
\begin{center}
\centerline{\epsfig{figure=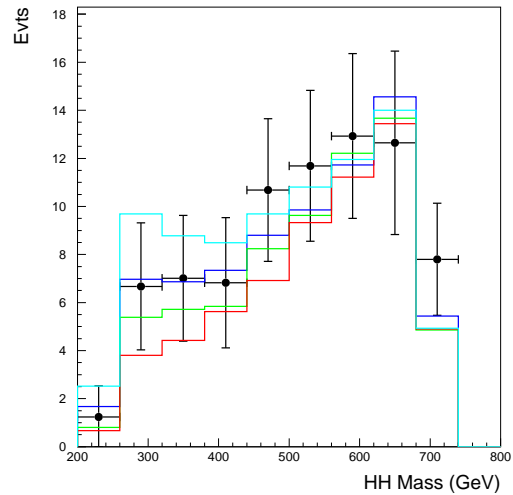,width=7.5cm}}
\caption{Invariant mass of the $HH$ system in $e^+e^- \to HHZ$ events reconstructed with 
1~ab$^{-1}$ of data at 0.8~TeV. The histograms show the predicted distribution for various
values of $g_{HHH}$ demonstrating that the low mass region is sensitive to the contribution 
of the triple Higgs vertex.}  
\label{fig9}%
\end{center}
\end{figure}
Other modes, such as $e^+e^- \to H H b \bar b$, have also been recently 
proposed~\cite{Gutierrez-Rodriguez:2006qk} but signal yields are too small to 
provide any precise data. Operating at $\sqrt{s}$ = 0.5~TeV the ILC can measure the $HHZ$ 
production cross section to about 15\% accuracy, if the Higgs boson mass is 120~GeV, 
corresponding to a fractional accuracy on $g_{HHH}$ of 23\%~\cite{Castanier:2001sf}. 
Improvements can be obtained first by introducing observables sensitive to the presence 
of the triple Higgs vertex and then by performing the analysis at higher 
energies where the $HH \nu \bar \nu$ channel contributes~\cite{Battaglia:2001nn}. 
In the $HHZ$ process events from diagrams containing the $HHH$ vertex exhibit a lower 
invariant mass of the $HH$ system compared to double-Higgstrahlung events. When the $M_{HH}$ 
spectrum is fitted, a relative statistical accuracy of $\pm 0.20$ can be obtained with 
1~ab$^{-1}$ at $\sqrt{s}$ = 0.5~TeV. 
The availability of beam polarization increases the $HHZ$ cross section by a factor of 
two and that for $HH \nu \bar \nu$ by a factor of four, thus offering a further possible 
significant improvement to the final accuracy. The ILC and, possibly, a multi-TeV 
$e^+e^-$ collider represent a unique opportunity for carrying out this fundamental 
measurement. In fact, preliminary studies show that, the analysis of double Higgs 
production at the LHC is only possible after a luminosity upgrade and, even then, 
beyond the observation of double Higgs production, it would provide only a very 
limited information on the triple-Higgs coupling ~\cite{Baur:2002qd,Baur:2003gp}.

\begin{table}[h]
\tbl{Summary of the accuracies on the determination of the Higgs boson 
profile at the ILC. Results are given for a 350-500~GeV ILC with 
${\cal{L}}$=0.5~ab$^{-1}$. Further improvements, expected from a 
1~TeV ILC are also shown for some of the measurements.}
{\begin{tabular}{@{}lcc@{}} \toprule
         & $M_H$            & $\delta(X)/X$ \\
         & (GeV)            & ILC-500~~~~ $|$ ILC-1000 \\
         &                  & 0.5~ab$^{-1}$ $|$ 1~ab$^{-1}$ \\ 
 \colrule
  $\delta M_H/M$      & 120-180     & (3-5) $\times 10^{-4}$ \\
  $\delta \Gamma_{tot}/\Gamma$ & 120-200 & 0.03- -- $|$  0.03 - 0.05 \\ \colrule
  $\delta g_{HWW}/g$ & 120-240 & 0.01-0.03 $|$ 0.01 - 0.01\\ 
  $\delta g_{HZZ}/g$ & 120-240 & 0.01-0.05 \\ \colrule
  $\delta g_{Htt}/g$ & 120-200 & 0.02- -- $|$ 0.06 - 0.13\\
  $\delta g_{Hbb}/g$ & 120-200 & 0.01-0.06 $|$ 0.01 - 0.05\\
  $\delta g_{Hcc}/g$ & 120-140 & 0.06-0.12 \\
  $\delta g_{H\tau\tau}/g$ & 120-140  & 0.03-0.05 \\ 
  $\delta g_{H\mu\mu}/g$ & 120-140  & 0.15 $|$ 0.04-0.06 \\ \colrule
  {\cal CP} test & 120   & 0.03  \\
  $\delta g_{HHH}/g$ & 120 & 0.20 $|$ 0.12 \\ \botrule
\end{tabular}
}
\label{tab:summary}
\end{table}

\subsection{Understanding Dark Matter at the ILC}\label{sec3.2}

The search for new physics beyond the Standard Model has a central role 
in the science program of future colliders.
It is instructive to contrast the LHC and the ILC in terms of their 
potential in such searches. Running at $\sqrt{s} \le$ 1~TeV the ILC might
appear to be limited in reach, somewhere within the energy domain being 
probed by the Tevatron and that to be accessed by LHC. And yet its potential 
for fully understanding the new physics, which the LHC might have manifested, 
and for probing the high energy 
frontier beyond the boundaries explored in hadron collisions is of paramount 
importance. There are several examples of how the ILC will be essential 
for understanding new physics. They address scenarios where signals of 
physics beyond the SM, as observed at the LHC, may be insufficient to 
decide on the nature of the new phenomena. One such example, which has 
been studied in some details, is the case of Supersymmetry and Universal 
Extra Dimensions (UED), two very different models of new physics leading 
to the very same experimental signature: fermion pairs plus missing 
energy. Here, the limited analytical power of the LHC may leave us 
undecided~\cite{Datta:2005zs,Smillie:2005ar}, while a single spin 
measurement performed at the ILC precisely identifies the nature of the 
observed particles~\cite{Battaglia:2005zf}. But the ILC capability to fully 
understand the implications of new physics, through fundamental measurements 
performed with high accuracy, is manifested also in scenarios where the 
LHC could observe a significant fraction of the new particle spectrum. 
An especially compelling example, which can be studied quantitatively, 
is offered by Supersymmetry in relation to Dark Matter (DM). Dark Matter has 
been established as a major component of the Universe.
We know from several independent observations, including the cosmic microwave 
background (CMB), supernovas (SNs) and galaxy clusters, that DM is responsible 
for approximately 20~\% of the energy density of the universe. Yet, none of the 
SM particles can be responsible for it and the observation of DM is likely the 
first direct signal of new physics beyond the SM. Several particles and objects
have been nominated as candidates for DM. They span a wide range of 
masses, from 10$^{-5}$~eV, in the case of axions, to 10$^{-5}$~solar masses, for 
primordial black holes. Cosmology tells us that a significant fraction of the 
Universe mass consists of DM, but does not provide clues on its nature. Particle 
physics tells us that New Physics must exist at, or just beyond, the EW scale 
and new symmetries may result in new, stable particles. Establishing the 
inter-relations between physics at the microscopic scale and phenomena at 
cosmological scale will represent a major theme for physics in the next decades. 
The ILC will be able to play a key role in elucidating these inter-relations.
Out of these many possibilities, there is a class of
models which is especially attractive since its existence is independently 
motivated and DM, at about the observed density, arises naturally. These are 
extensions of the SM, which include an extra symmetry protecting the lightest 
particle in the new sector from decaying into ordinary SM states. 
The lightest particle becomes stable and can be chosen to be 
neutral. Such a particle is 
called a weakly interacting massive particle (WIMP) and arises in 
Supersymmetry with conserved R-parity (SUSY) but also in Extra Dimensions 
with KK-parity (UED)~\cite{Kong:2005hn}. Current cosmological data, mostly 
through the WMAP satellite measurements of the CMB, determine the DM density 
in the Universe with a 6~\% relative accuracy~\cite{wmap}. By the next decade, 
the PLANCK satellite will push this uncertainty to $\simeq$~1~\%, or 
below~\cite{planck}. Additional astrophysical data manifest a possible 
evidence of DM annihilation. The EGRET data show excess of $\gamma$ emission 
in the inner galaxy, which has been interpreted as due to DM~\cite{deBoer:2005tm}
and the WMAP data itself may show a signal of synchrotron emission in the 
Galactic center~\cite{Finkbeiner:2004us}. These data, if confirmed, may be used to 
further constrain the DM properties. Ground-based DM searches are 
also approaching the stage where their sensitivity is at the level predicted by 
Supersymmetry for some combinations of parameters~\cite{Akerib:2005kh}.
The next decades promise to 
be a time when accelerator experiments will provide new breakthroughs and highly 
accurate data to gain new insights, not only on fundamental questions in particle 
physics, but also in cosmology, when studied alongside the observations from 
satellites and other experiments. The questions on the nature and the origin of 
DM offer a prime example of the synergies of new experiments at hadron and lepton 
colliders, at satellites and ground-based DM experiments.

It is essential to study, in well defined, yet general enough, models, 
which are the properties of the new physics sector, such as masses and couplings,
most important to determine the resulting relic density of the DM particles. 
Models exist which allow to link the microscopic particle properties to the present 
DM density in the Universe, with mild assumptions. If DM consists of WIMPs, they are 
abundantly produced in the very early Universe when 
$T \simeq (t \mathrm{(sec)})^{-1/2} > $~100~GeV and their interaction cross section is 
large enough that they were in thermal equilibrium for some period in the early universe. 
The DM relic density can be determined by solving the Boltzmann equation governing the 
evolution of their phase space number density~\cite{Scherrer:1985zt}. 
It can be shown that, by taking the WMAP result for the DM relic density in units of the 
Universe critical density,$\Omega_{DM} h^2$, the thermal averaged DM annihilation 
cross section times the co-moving velocity, $<\sigma v>$, should be $\simeq$~0.9. From 
this result, the mass of the DM candidate can be estimated as:
\begin{equation}
M_{DM}  = \sqrt{\frac{\pi \alpha^2}{8 <\sigma v>}} \simeq 100~\mathrm{GeV}.
\end{equation}
A particle with mass $M = {\cal{O}}$(100~GeV) and weak cross section would naturally 
give the measured DM density. It is quite suggestive that new physics, responsible for 
the breaking of electro-weak symmetry, also introduce a WIMP of about that mass. 
In fact, in essentially every model of electroweak symmetry breaking, it is possible 
to add a discrete symmetry that makes the lightest new particle stable. 
Often, this discrete symmetry is required for other reasons. 
For example, in Supersymmetry, the conserved $R$ parity is needed to eliminate 
rapid proton decay.  In other cases, such as models with TeV-scale extra
dimensions, the discrete symmetry is a natural consequence of the 
underlying geometry.  

\begin{figure}
\centerline{\psfig{file=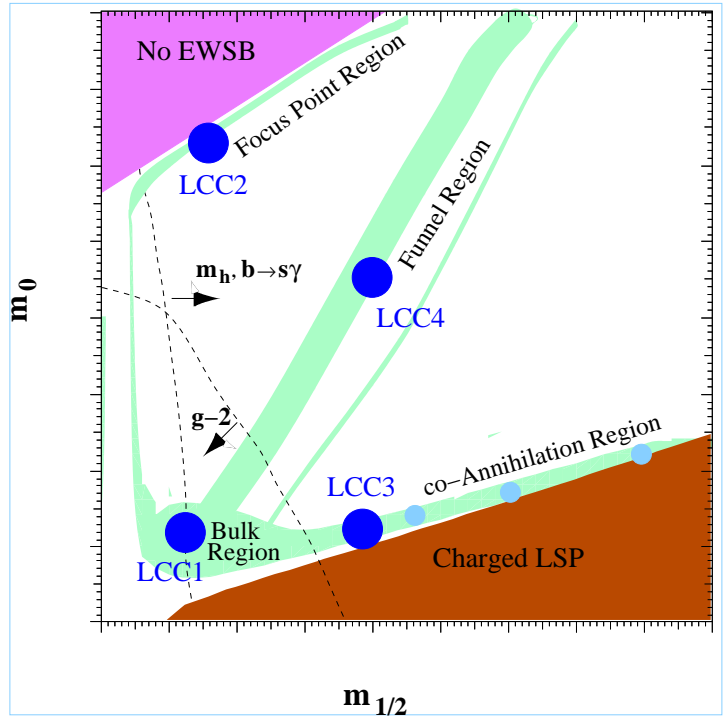,width=9.5cm}}
\caption{The DM-favoured regions in the $m_0$ - $m_{1/2}$ plane of the cMSSM and 
existing constraints. The precise locations of these regions vary with the $\tan \beta$ 
parameter and therefore the axis are given without units. The indicative locations of 
the four benchmark points adopted, are also given. Lower limits on the Higgs boson mass 
and, in a portion of the parameter space, the measurement of the $b \to s \gamma$ 
decay branching fraction, exclude the region at low values of $m_{1/2}$. 
A discrepancy of the measured anomalous magnetic moment of the muon value with the 
SM prediction would favour the region on the left of the curve labeled $g-2$.}
\label{fig10}
\end{figure}

Data on DM density already set rather stringent constraints on the parameters
of Supersymmetry, if the lightest neutralino $\chi^0_1$ 
is indeed responsible for saturating the amount of DM observed in the Universe. 
It is useful to discuss the different scenarios, where neutralino DM density is 
compatible with the WMAP result, in terms of parameter choices in the context 
of the constrained MSSM (cMSSM), to understand how the measurements that the ILC 
provides can establish the relation between new physics and DM.  
The cMSSM reduces the number of free parameters to just five: the common 
scalar mass, $m_0$, the common gaugino mass, $m_{1/2}$, the ratio of the vacuum 
expectation values of the two Higgs fields, $\tan\beta$, the sign of the Higgsino 
mass parameter, $\mu$, and the common trilinear coupling, $A_0$. It is a remarkable 
feature of this model that, as these parameters, defined at the unification scale, 
are evolved down to lower energies, the electroweak symmetry is broken spontaneously 
and masses for the $W^{\pm}$ and $Z^0$ bosons generated automatically. As this model 
is simple and defined by a small number of parameters, it is well suited for 
phenomenological studies. The cosmologically interesting regions in the 
$m_0$ - $m_{1/2}$ parameter plane are shown in Figure~\ref{fig10}. 
As we move away from the bulk region, at small values of $m_0$ and $m_{1/2}$, which is 
already severely constrained by LEP-2 data, the masses of supersymmetric particles 
increase and so does the dark matter density. It is therefore necessary to have an 
annihilation process, which could efficiently remove neutralinos in the early universe, 
to restore the DM density to the value measured by WMAP.
Different processes define three main regions: i) the focus point region, 
where the $\chi^0_1$ contains an admixture of the supersymmetric partner of a neutral 
Higgs boson and annihilates to $W^+W^-$ and $Z^0Z^0$, ii) the co-annihilation region, 
where the lightest slepton has a mass very close to $M_{\chi^0_1}$,  
iii) the $A$ annihilation funnel, where $M(\chi^0_1)$ is approximately half 
that of the heavy $A^0$ Higgs boson, providing efficient $s$-channel 
annihilation, $\chi \chi \to A$. In each of these regions, researchers at the 
ILC will be confronted with several different measurements and significantly 
different event signatures. 

\begin{table}[h]
\tbl{cMSSM parameters of benchmark points}
{\begin{tabular}{@{}lcccccc@{}} \toprule
Point & $m_0$ &  $m_{1/2}$ &  $\tan\beta$ &  $A_0$ & $Sgn(\mu)$ & 
$M(t)$ \\ \colrule
LCC1 &  100  &  250  &  10   &  -100 & + & 178 \\
LCC2 &  3280 &  300  &  10   &  0    & + & 175 \\ 
LCC3 &  210  &  360  &  40   &  0    & + & 178 \\ 
LCC4 &  380  &  420  &  53   &  0    & + & 178 \\ \botrule
\end{tabular}}
\label{tab:bench}
\end{table}

It is interesting to observe that the DM constraint, reduces the dimensionality 
of the cMSSM plane, by one unit, since the allowed regions are tiny lines in 
the $m_0$ - $m_{1/2}$ plane, evolve with $\tan \beta$ and depend only very weakly 
on $A_0$~\cite{Battaglia:2003ab}. Representative benchmark points have been defined 
and their parameters are summarised in Table~\ref{tab:bench}. Even though these 
points have been defined in a specific supersymmetric model, their phenomenology 
is common to the more general supersymmetric solutions and we shall soon discuss 
the extension of results derived in this constrained model to the general MSSM.
There are several features which are common to all these regions. First,
the relic density depends on the mass of the lightest neutralino and of few 
additional particles, close in mass to it. The heavier part of the SUSY spectrum 
decouples from the value of $\Omega_{\chi} h^2$. This is of particular importance 
for the ILC. Running at $\sqrt{s} \le$ 1~TeV, the ILC will not be able to study 
supersymmetric particles exceeding $\simeq$450-490~GeV, in particular scalar 
quarks and heavy Higgs bosons in some regions of the parameter phase space. But,
independently of the LHC results, the ILC will either observe and measure these 
particles if they may be relevant to determine the relic DM density, or it will 
set bounds that ensure their decoupling. A second important observation is 
that $\Omega_{\chi} h^2$ typically depends on SUSY parameters which can be fixed 
by accurate measurements of particle masses, particle mass splittings, decay 
branching fractions and production cross sections. In some instances the 
availability of polarised beams is advantageous. The LHC can often make precise
measurements of some particles, but it is difficult for the LHC experiments to 
assemble the complete set of parameters needed to
reconstruct annihilation cross section.  It is also typical of supersymmetry
spectra to contain light particles that may be very difficult to observe in the
hadron collider environment.  The ILC, in contrast, provides just the right
setting to obtain both types of measurements.  Again, it is not necessary
for the ILC to match the energy of the LHC, only that it provides enough
energy to see the lightest charged particles of the new sector.
 
Rather detailed ILC analyses of the relevant channels for each benchmark point have 
been performed,~\cite{Weiglein:2004hn,Gray:2005ci,Khotilovich:2005gb,Battaglia:2004gk} 
based on parametric simulation, which includes realistic detector 
performances and effects of the ILC beam characteristics. It has been assumed 
that the ILC will be able to provide collisions at centre-of-mass energies 
from 0.3~TeV to 0.5~TeV with an integrated luminosity of 500~fb$^{-1}$ in a first 
phase of operation and then its collision energy can raised 
to 1~TeV to provide an additional data set of 1~ab$^{-1}$, corresponding to an 
additional three to four years of running. Results are summarised in terms of the 
estimated accuracies on masses and mass differences in Table~\ref{tab:constraints}.

\begin{table}[h]
\tbl{Summary of the accuracies (in GeV) on the main mass determinations by the 
ILC at 0.5~TeV for the four benchmark points. Results in [] brackets also include 
ILC data at 1~TeV.}
{\begin{tabular}{@{}lcccc@{}} \toprule
Observable & LCC1 & LCC2 & LCC3 & LCC4 \\ \colrule
$\delta M(\tilde \chi^0_1)$ & $\pm$ 0.05 & $\pm$ 1.0 & $\pm$ 0.1 & [$\pm$ 1.4] \\
$\delta M(\tilde e_R) $ &  $\pm$ 0.05 & - & [$\pm$ 1.0] & [$\pm$ 0.6] \\
$\delta M(\tilde \tau_1)$ &   $\pm$ 0.3 & - & $\pm$ 0.5 & $\pm$ 0.9  \\
$\delta M(\tilde \tau_2)$ &  $\pm$ 1.1 & - & - & - \\
$\delta (M(\mu_R) - M(\tilde \chi^0_1))$ & $\pm$ 0.2 &  & [$\pm 0.2$] & $\pm$ 0.6 \\
$\delta (M(\tilde \tau_1) - M(\tilde \chi^0_1))$ & 0.3 & - & $\pm$ 1.0 & $\pm$ 1.0 \\
$\delta (M(\tilde \tau_2) - M(\tilde \chi^0_1))$ & $\pm$ 1.1 &  & [$\pm$ 3.0] & \\
$\delta (M(\tilde\chi^0_2)-M(\tilde\chi^0_1))$ & $\pm$ 0.07 & $\pm$ 0.3 & $\pm$ 0.6 & [$\pm$ 1.8]\\
$\delta (M(\tilde\chi^0_3)-M(\tilde\chi^0_1))$ & $\pm$ 4.0 & $\pm$ 0.2 & [$\pm$ 2.0] & [$\pm$ 2.0] \\
$\delta (M(\tilde \chi^+_1)-M(\tilde \chi^0_1))$ & $\pm$ 0.6  & $\pm$ 0.25 & [$\pm$ 0.7] & $\pm$ 2.0 \\
$\delta (M(\tilde \chi^+_2)-M(\tilde \chi^+_1))$ & [$\pm$ 3.0] & - & [$\pm$ 2.0] & $\pm$ 2.0 \\
$\delta M(A^0) $ & [$\pm$ 1.5] & - &  [$\pm$ 0.8] & [$\pm$ 0.8] \\
$\delta \Gamma(A^0) $ &  & - & [$\pm$ 1.2] & [$\pm$ 1.2] \\ \botrule
\end{tabular}}
\label{tab:constraints}
\end{table}

In order to estimate the implications of these ILC measurements
on the estimation of neutralino dark DM density $\Omega_{\chi} h^2$, broad scans 
of the multi-parameter supersymmetric phase space need to be performed. For each 
benchmark point, the soft parameters (masses and couplings) at the electroweak 
scale can be computed with the full 2-loop renormalization group equations 
and threshold corrections using {\tt Isajet 7.69}~\cite{ISAJET}.
Supersymmetric loop corrections to the Yukawa couplings can also be
included. The electroweak-scale MSSM parameters are extracted from the high 
scale cMSSM parameters. The dark matter density $\Omega_{\chi} h^2$ can be 
estimated using the {\tt DarkSUSY}~\cite{Gondolo:2004sc} and 
{\tt Micromegas}~\cite{Belanger:2006is} programs. 
These programs use the same {\tt Isajet} code to determine the particle 
spectrum and couplings, including the running Yukawa couplings, and 
compute the thermally averaged cross section for neutralino annihilation, 
including co-annihilation and solve the equation describing the evolution of 
the number density for the DM candidate. While the assumptions of the cMSSM are 
quite helpful for defining a set of benchmark points, the cMSSM is not 
representative of the generic MSSM, since it implies several mass relations, 
and its assumptions have no strong physics justification. Therefore, in 
studying the accuracy on $\Omega_{\chi} h^2$, the full set of MSSM parameters 
must be scanned in an uncorrelated way and the mass spectrum evaluated for 
each parameter set. A detailed study has recently been performed~\cite{Baltz:2006fm}. 
I summarise here some of the findings, Table~\ref{tab:dmsummary} gives 
results for the neutralino relic density estimates in MSSM for the LHC, the 
ILC at 0.5~TeV and the ILC at 1~TeV.  

The LCC1 point is in the bulk region and the model contains light sleptons, with 
masses just above that of the lightest neutralino. The most important annihilation
reactions are those with t-channel slepton exchange. At the LHC, many of the SUSY 
spectrum parameters can be determined from kinematic constraints. At the ILC masses 
can be determined both by the two-body decay kinematics of the pair-produced 
SUSY particles and by dedicated threshold scans. Let us consider the two body decay
of a scalar quark $\tilde q \to q \chi^0_1$. If the scalar quarks are pair produced 
$e^+e^- \to \tilde q \tilde q$, $E_{\tilde q} = E_{beam}$ and the $\chi^0_1$ escapes 
undetected,  only the $q$ (and the $\bar q$) are observed in the detector. In a 1994 
paper, J.~Feng and D.~Finnell~\cite{Feng:1993sd} pointed out that the minimum and 
maximum energy of production for the quark can be related to the mass difference
between the scalar quark $\tilde q$ and the $\chi^0_1$:

\begin{equation}
E_{max,~min} = \frac{E_{beam}}{2} \big( 1 \pm \sqrt{1 - \frac{m_{\tilde q}^2}{E_{beam}^2}}
\big) \big( 1 - \frac{m_{\chi}^2}{m_{\tilde q}^2} \big).
\end{equation} 

The method can also be extended to slepton decays $\tilde \ell \to \ell \chi^0_1$, which 
share the same topology, and allows to determine slepton mass once that of the neutralino 
is known or determine a relation between the masses and get $m_{\chi^0_1}$ if that of the 
slepton can be independently measured. The measurement requires a precise determination 
of the endpoint energies of the lepton momentum spectrum, $E_{min}$ and $E_{max}$.
It can be shown that accuracy is limited by beamstrahlung, affecting the knowledge of 
$E_{beam}$ in the equation above, more than by the finite momentum resolution, 
$\delta p/p$ of the detector. 
\begin{figure}[h]
\begin{center}
  \minifig{2.1in}{\epsfig{figure=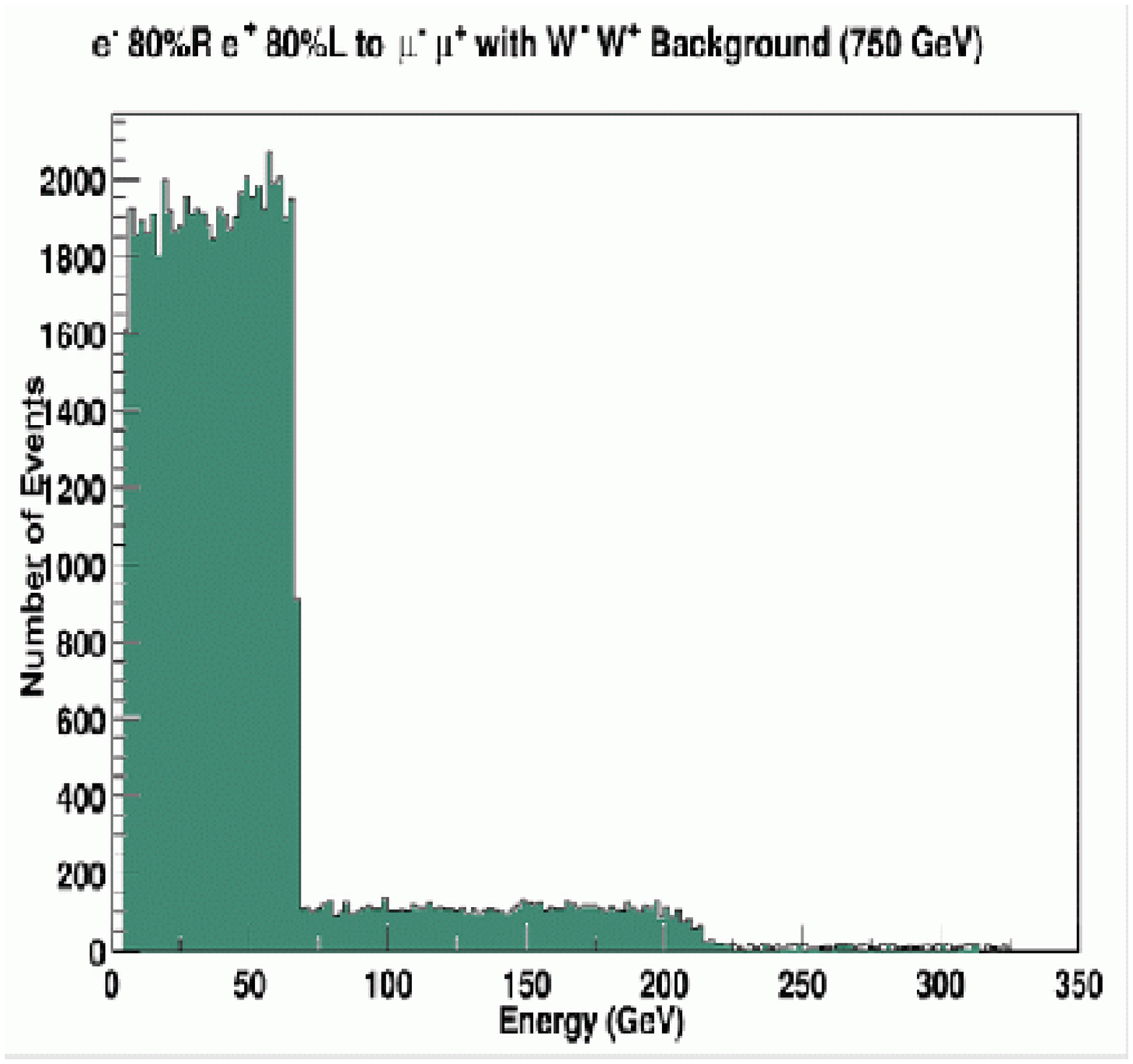,height=5.0cm,clip}}{a}
  \hspace*{15pt}
  \minifig{2.1in}{\epsfig{figure=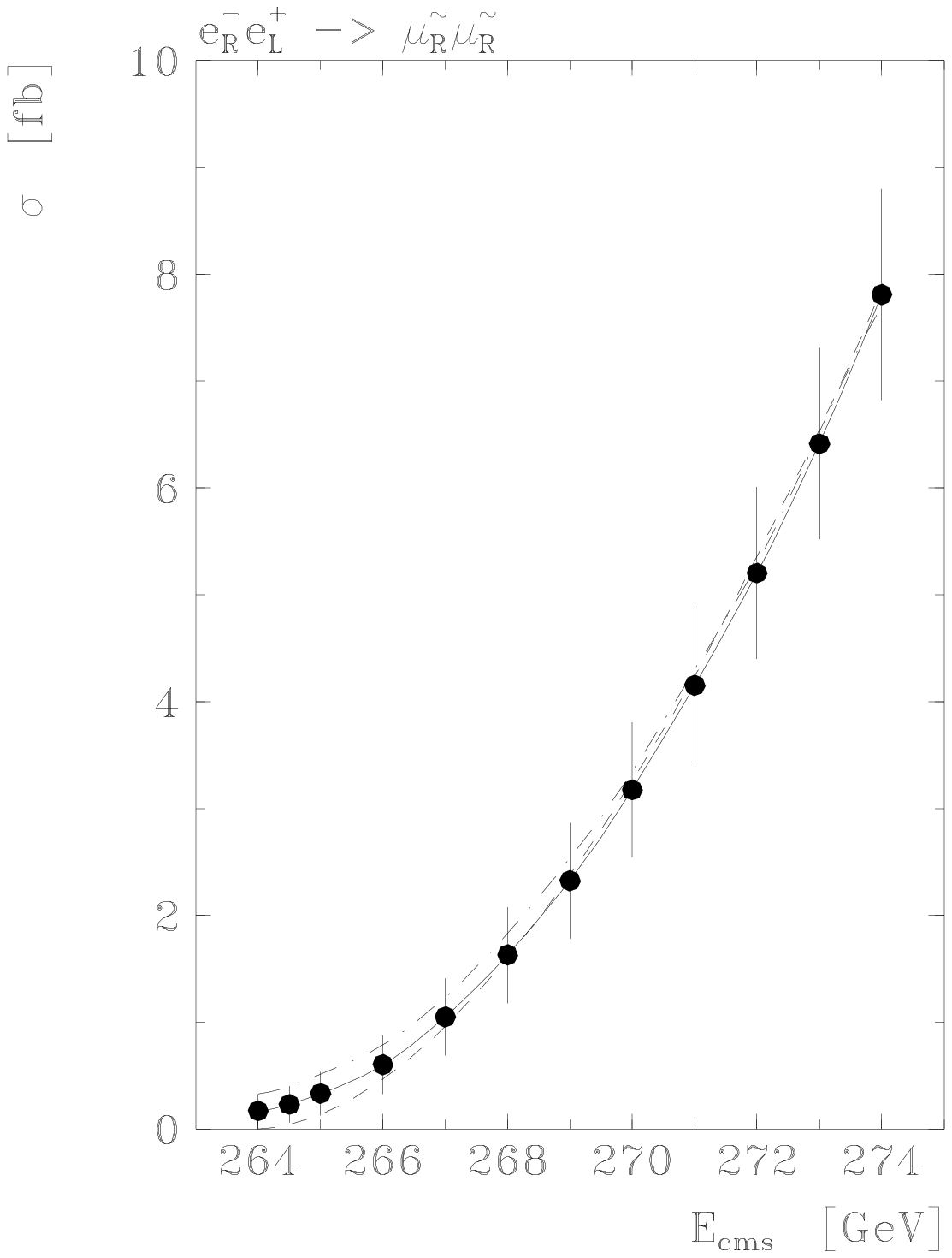,height=5.75cm,clip}}{b}
\caption{Mass reconstruction at ILC: (a) momentum endpoint in $\tilde \mu \to \mu \chi^0_1$ 
(from Ref.~\cite{Moortgat-Pick:2005cw}) and (b) threshold scan for 
$e^+e^ \to \tilde \mu^+ \tilde \mu^-$ (from Ref.~\cite{Aguilar-Saavedra:2001rg})} 
\label{fig11}%
\end{center}
\end{figure}
The ILC has a second, and even more precise, method for mass measurements.
The possibility to precisely tune the collision energy allows to perform 
scans of the onset of the cross section for a specific SUSY particle pair production 
process. The particle mass and width can be extracted from a fit to the signal event 
yield as function of $\sqrt{s}$. The accuracy depends rather weakly on the number of 
points, $N$, adopted in the scan and it appears that concentrating the total luminosity 
at two or three different energies close to the threshold is 
optimal~\cite{Martyn:2000,Blair:2001cz}. The mass accuracy, $\delta m$  can be 
parametrised as:

\begin{equation}
\delta m \simeq \Delta E \frac{1+0.36/\sqrt{N}}{\sqrt{18 N L \sigma}}
\end{equation}
for S-wave processes, where the cross section rises as $\beta$ and as

\begin{equation}
\delta m \simeq \Delta E \frac{1}{N^{1/4}} \frac{1+0.38/\sqrt{N}}{\sqrt{2.6 N L \sigma}} 
\end{equation}
for P-wave processes, where the cross section rises as $\beta^3$.
The combination of these measurements allows the ILC to determine the $\chi^0_1$ 
mass to $\pm$0.05~GeV, which is two orders of magnitude better than the anticipated 
LHC accuracy, while the mass difference between the $\tilde \tau_1$ and the $\chi^0_1$ 
can be measured to $\pm$0.3~GeV, which is more than a factor ten better. Extension of 
ILC operation to 1~TeV gives access to the $e^+e^- \to H^0A^0$ process. As a result of 
the precision of these measurements, the ILC data at 0.5~TeV will allow to predict the 
neutralino relic density to $\pm$2~\% and the addition of 1.0~TeV data will improve it 
to $\pm$0.25~\%. It is suggestive that this accuracy is comparable, or better, than that 
expected by the improved CMB survey by the PLANCK mission. For comparison, the LHC data 
should provide a $\pm$7~\% accuracy. This already a remarkable result, due the fact that, 
a large number of measurements will be available at the LHC and SUSY decay chains can be 
reconstructed. Still, the overall mass scale remains uncertain at the LHC. The direct mass 
measurements on the ILC data remove this uncertainty. 
\begin{figure}[h]
\begin{center}
\epsfig{figure=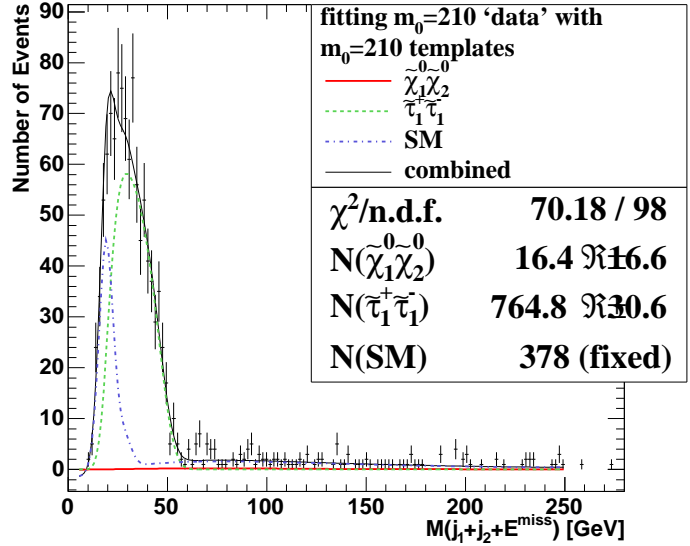,height=7.5cm,clip}
\caption{DM-motivated SUSY $\tilde \tau$ reconstruction at ILC: determination of the 
stau-neutralino mass difference from a reconstruction of 
$e^+e^- \to \tilde \tau_1 \tilde \tau_1$ at 0.5~TeV for LCC3 
(from Ref.~\cite{Khotilovich:2005gb}).} 
\label{fig11-2}%
\end{center}
\end{figure}

\begin{figure}[h]
\begin{center}
  \minifig{2.1in}{\epsfig{figure=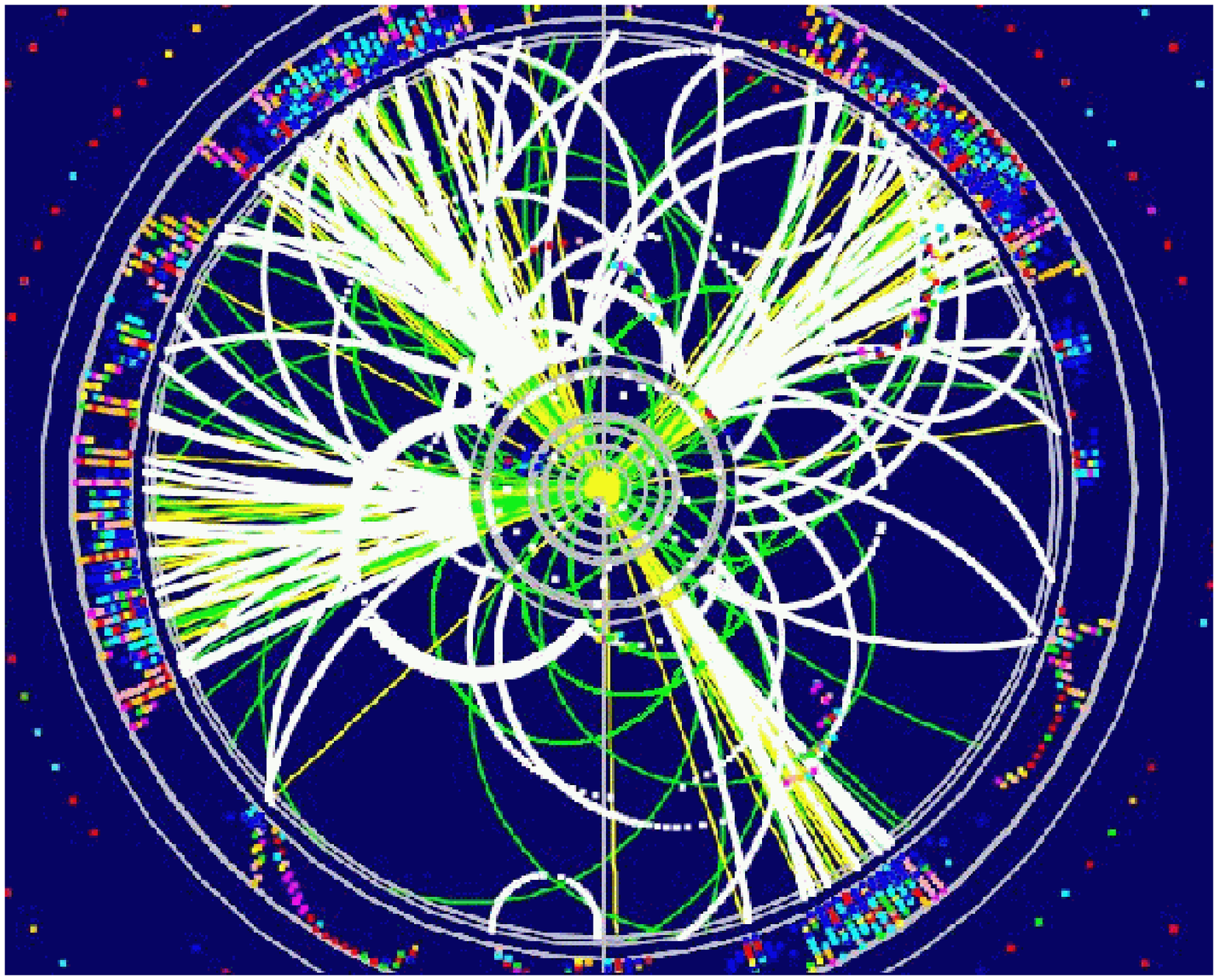,height=4.0cm,clip}}{a}
  \hspace*{1pt}
  \minifig{2.1in}{\epsfig{figure=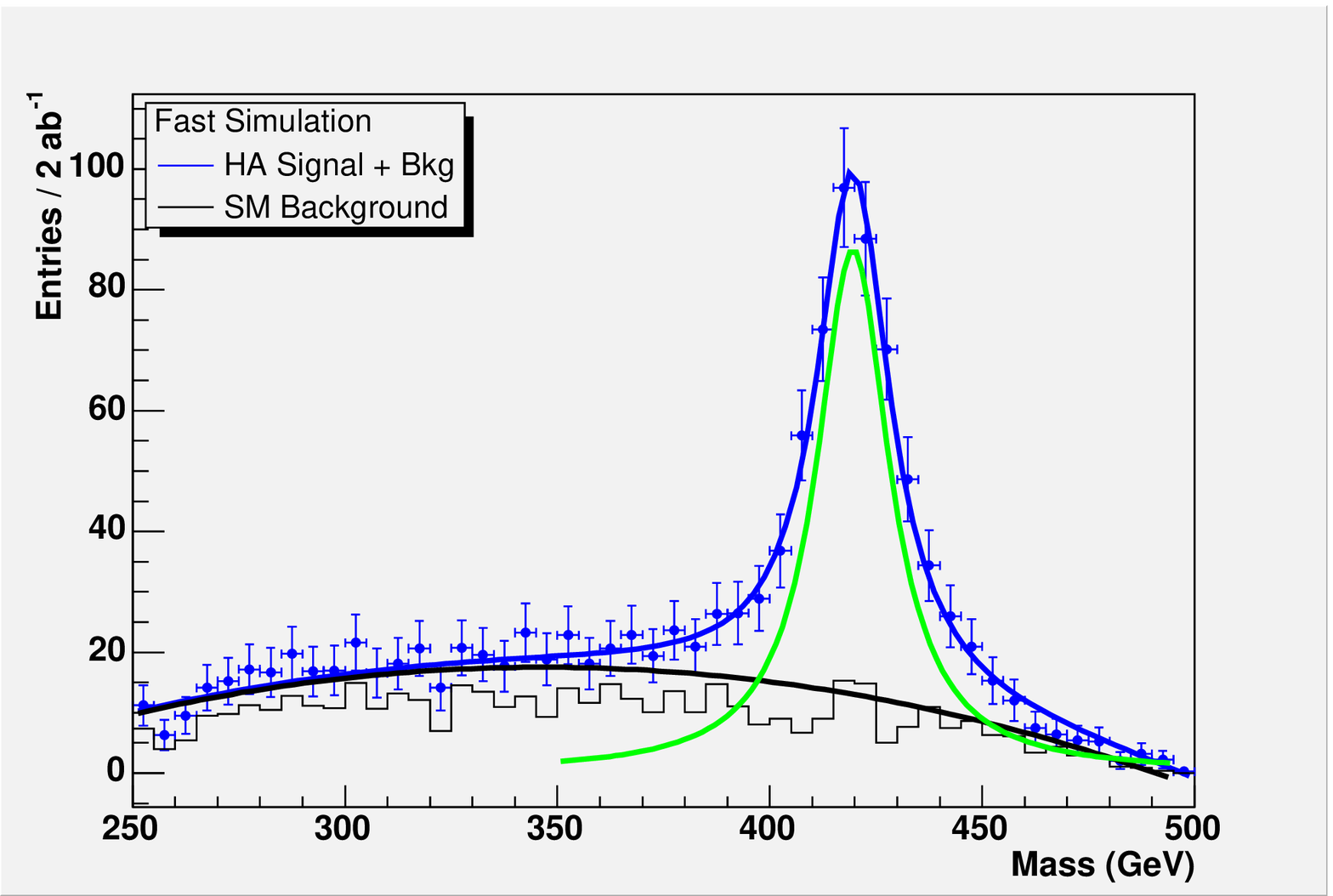,height=4.0cm,clip}}{b}
\caption{DM-motivated SUSY Higgs reconstruction at ILC: (a) an event 
$e^+e^- \to A^0 H^0 \to b \bar b b \bar b$ at 1~TeV in the LDC detector and 
(b) di-jet invariant mass spectrum for 
$e^+e^- \to A^0 H^0 \to b \bar b b \bar b$ at 1~TeV for LCC4 (from Ref.~\cite{Battaglia:2004gk}).} 
\label{fig12}%
\end{center}
\end{figure}
The LCC1 point is characterised by the relatively low SUSY mass scale, most of the 
particles can be observed at the LHC and their masses accurately measured at the ILC. 
However, in more general scenarios, the information available from both collider 
will be more limited. This is the case at the LCC2 point, located in the focus point 
region, where masses of scalar quarks, sleptons and heavy Higgs bosons are very large, 
typically beyond the ILC but also the LHC reach, while gauginos masses are of the order 
of few hundreds GeV, thus within the kinematical domain of the ILC. 
In this specific scenario, the LHC will observe the SUSY 
process $\tilde g \to q \bar q \chi$ and the subsequent neutralino and chargino decays. 
Still the neutralino relic density can only be constrained within $\pm$40\% and the 
hypothesis $\Omega_{\chi} h^2$ = 0, namely that the neutralino does not contribute to the 
observed dark matter density in the universe, cannot be ruled out, based only on LHC data.
At a 0.5~TeV collider, the main SUSY reactions are $e^+e^- \to \chi^+_1 \chi^-_1$ and 
$e^+e^- \to \chi^0_2 \chi^0_3$. Operation at 1~TeV gives access also to 
$e^+e^- \to \chi^+_2 \chi^-_2$ and $e^+e^- \to \chi^0_3 \chi^0_4$. Not only the gaugino mass
splittings but also the polarised neutralino and chargino production cross section 
can be accurately determined at the ILC~\cite{Gray:2005ci}.
These measurements fix the gaugino-Higgsino mixing angles, 
which play a major role in determining the neutralino relic density. The decoupling of the 
heavier, inaccessible part of the SUSY spectrum, can be insured with the data at the highest 
energy. The combined ILC data at 0.5~TeV and 1~TeV provide an estimate of the neutralino
relic density to $\pm$8~\% accuracy, which matches the current WMAP precision. 
The characteristics featured by the LCC2 point persist, while the SUSY masses increase, 
provided the gaugino-Higgsino mixing angle remains large enough. This DM-motivated region 
extends to SUSY masses which eventually exceed the LHC reach, highlighting an intriguing 
region of parameters where the ILC can still observe sizable production of supersymmetric 
particle, compatible with dark matter data, while the LHC may report no signals of 
New Physics~\cite{Baer:2003ru}.

Instead, the last two points considered, LCC3 and LCC4, are representative of those 
regions where the neutralino relic density is determined by accidental relationships 
between particle masses. Other such regions may also be motivated by 
baryogenesis constraints~\cite{Balazs:2004bu}.
The determination of the neutralino relic density, in such scenarios, depends crucially 
on the precision of spectroscopic measurements, due to the large sensitivity on masses 
and couplings. The conclusions of the current studies are that the LHC data do not provide 
quantitative constraints. On the contrary, the ILC can obtain interesting precision, especially 
when high energy data is available. 
\begin{figure}[h]
\begin{center}
  \minifig{2.1in}{\epsfig{figure=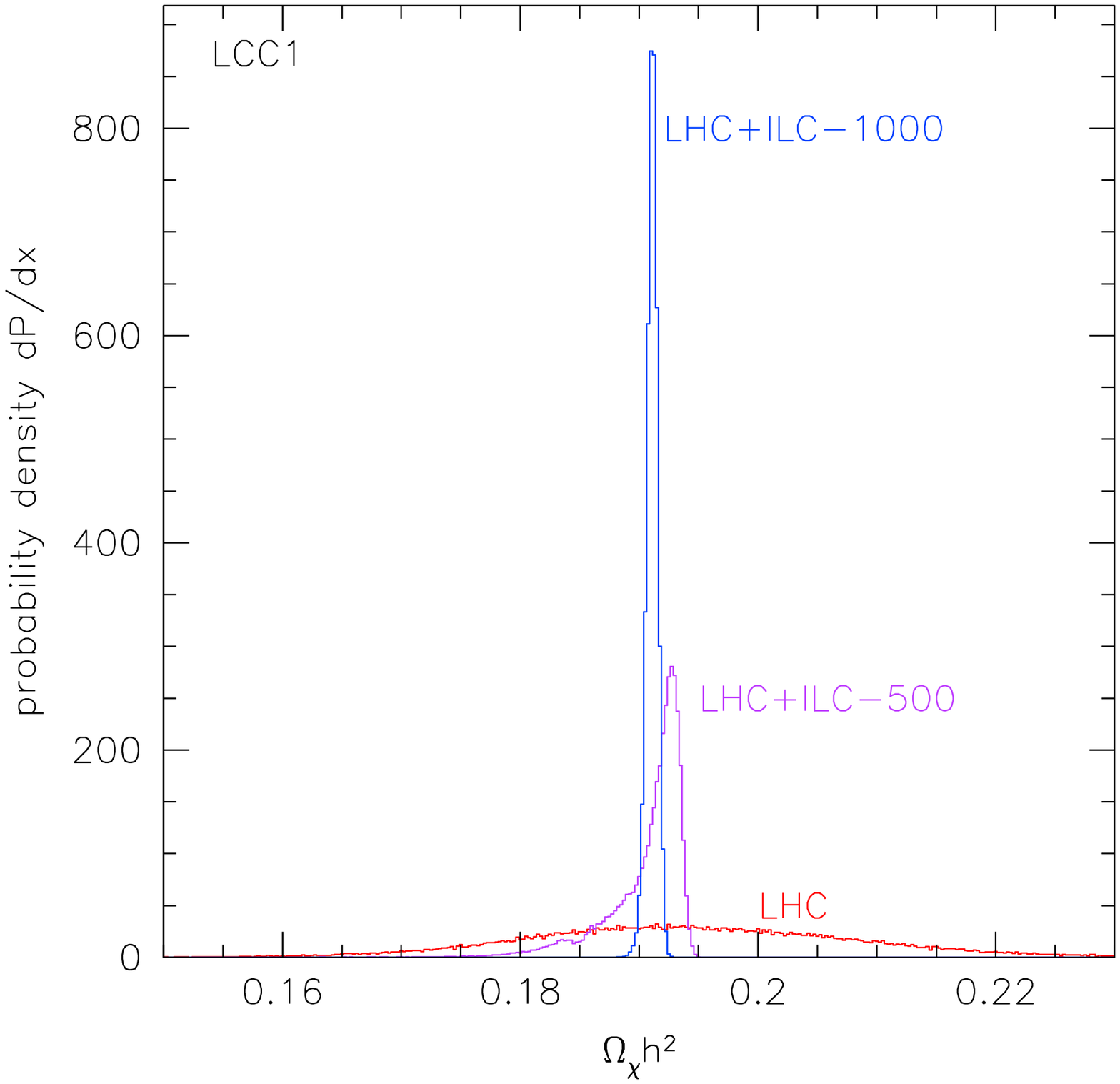,width=5.0cm}}{a}
  \hspace*{2pt}
  \minifig{2.1in}{\epsfig{figure=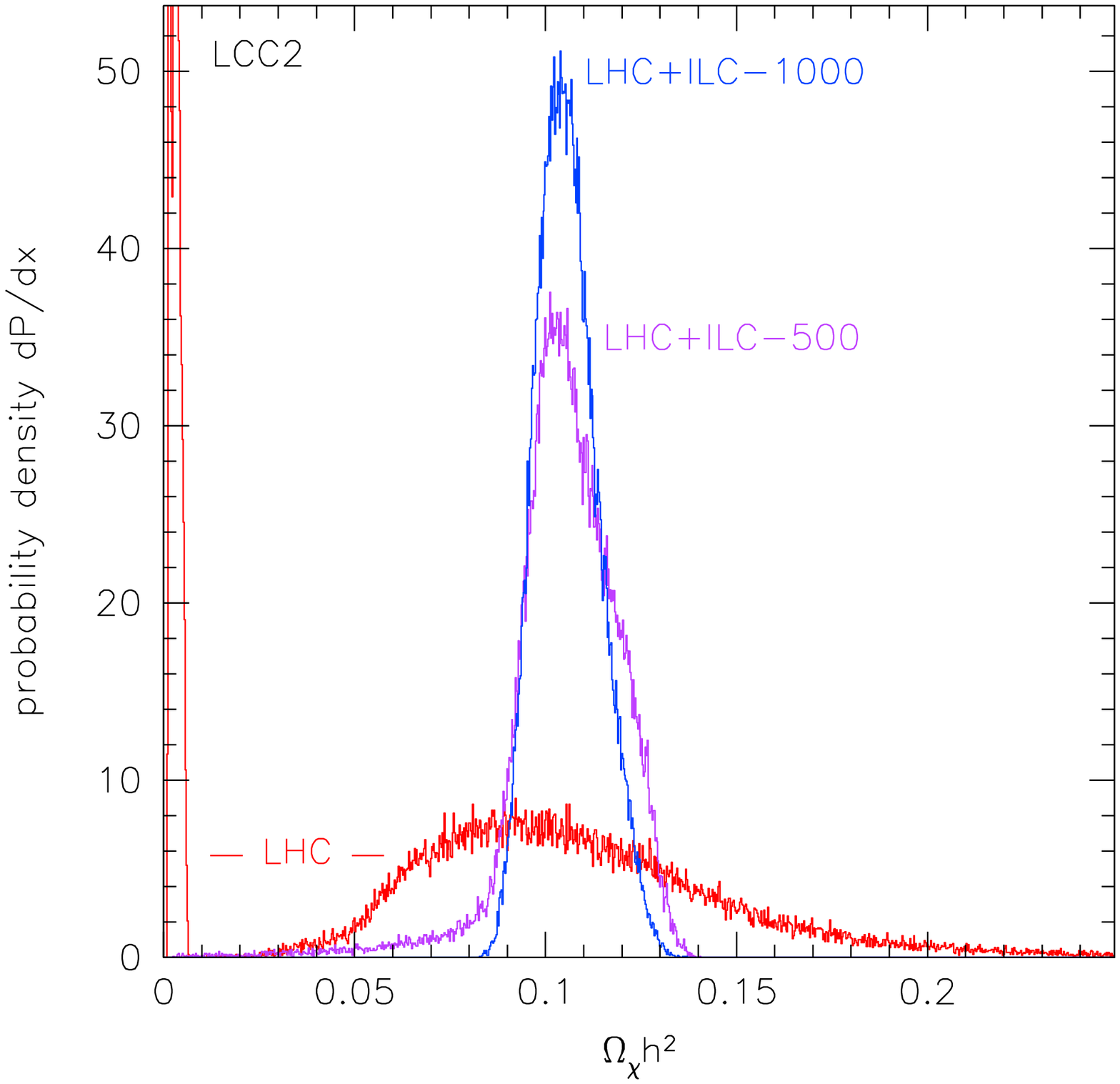,width=5.0cm}}{b}
  \vspace*{2pt}
  \minifig{2.1in}{\epsfig{figure=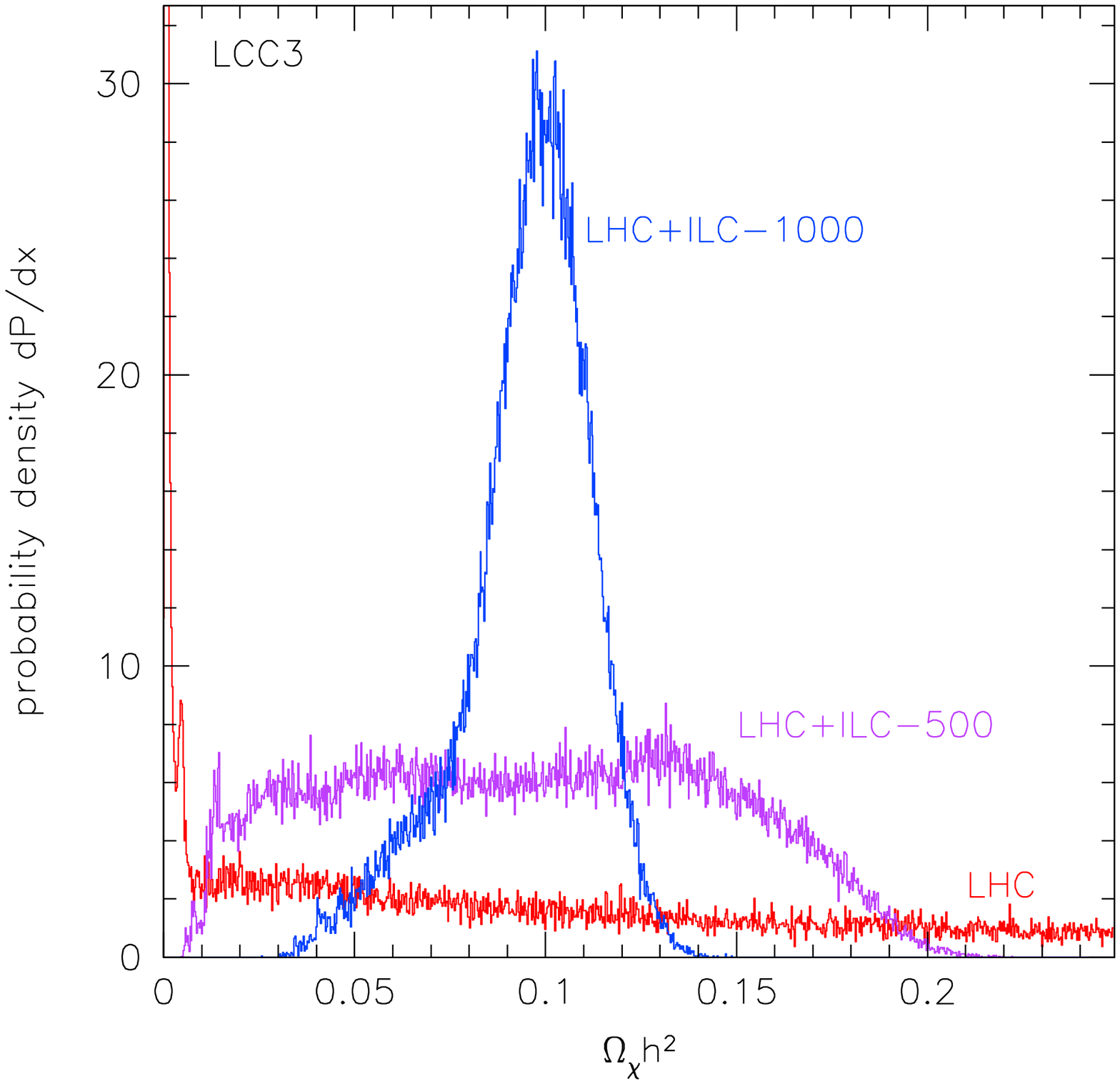,width=5.0cm}}{c}
  \hspace*{2pt}
  \minifig{2.1in}{\epsfig{figure=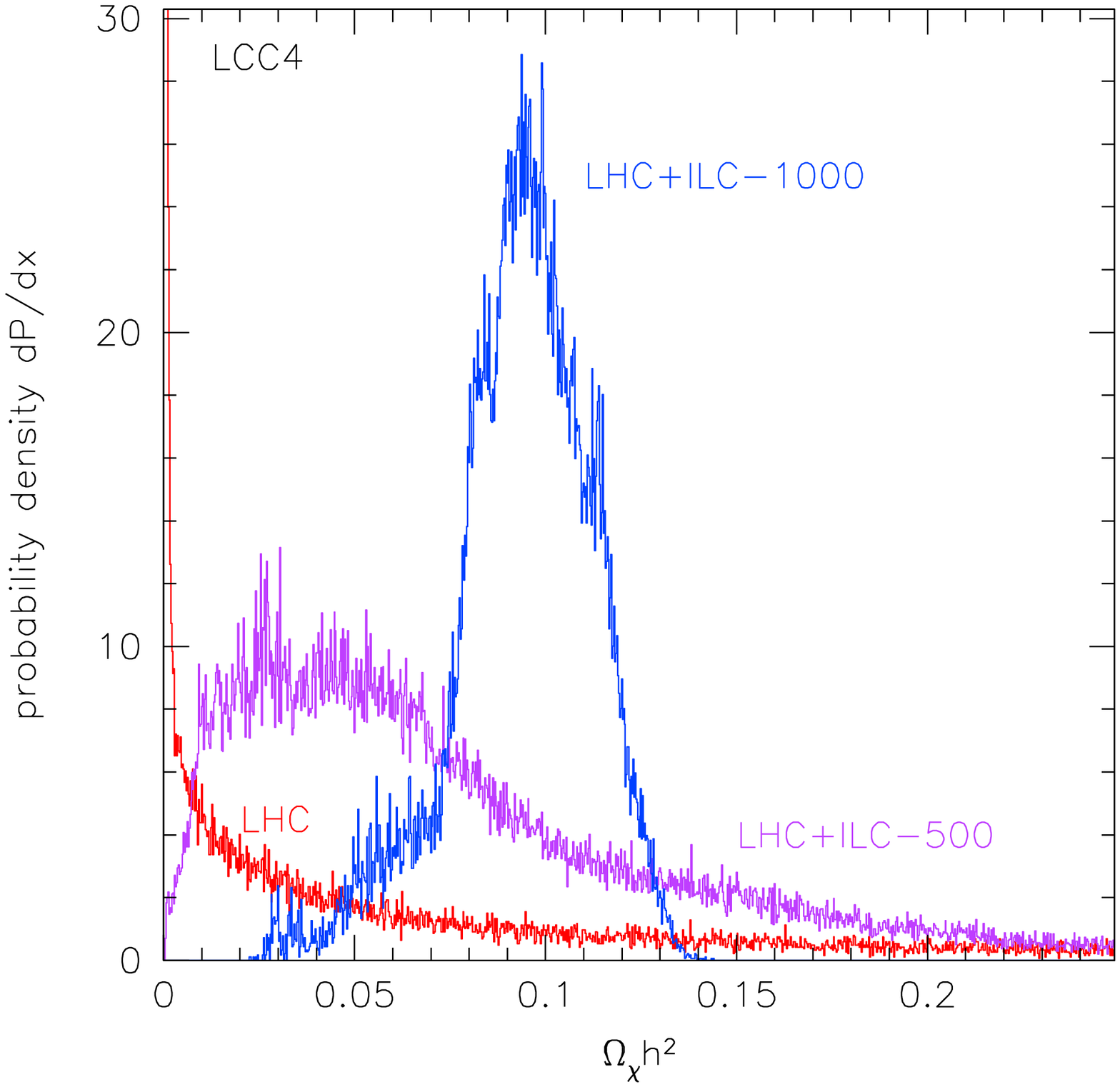,width=5.0cm}}{d}
\caption{Relic DM density determination based on simulation from LHC, ILC at 0.5~TeV and 
ILC at 1.0~TeV for the four SUSY benchmark points studied: a) LCC1, b) LCC2, c) LCC3 and d) LCC4. 
The plots show the probability density functions of the $\Omega h^2$ values corresponding to MSSM 
points compatible with the accelerator data (from Ref.~\cite{Baltz:2006fm}).}
  \label{fig13}%
\end{center}
\end{figure}
\begin{table}[h]
\tbl{Summary of the relative accuracy $\frac{\delta \Omega_{\chi} h^2}{\Omega_{\chi} h^2}$ for the
four benchmark points obtained with full SUSY scans.}
{\begin{tabular}{@{}lcccc@{}} \toprule
Benchmark & $\Omega h^2$ & LHC & ILC & ILC \\
Point     &              &     & 0.5~TeV & 1.0~TeV \\ \colrule
LCC1      & 0.192        & 0.072 & 0.018 & 0.024 \\
LCC2      & 0.109        & 0.820 & 0.140 & 0.076 \\
LCC3      & 0.101        & 1.670 & 0.500 & 0.180 \\
LCC4      & 0.114        & 4.050 & 0.850 & 0.190 \\
\botrule
\end{tabular}}
\label{tab:dmsummary}
\end{table}
The LCC3 point is in the so-called $\tilde \tau$ co-annihilation 
region. Here, the mass difference between the lightest neutralino, $\chi^0_1$, and 
the lightest scalar tau, $\tilde \tau_1$, is small enough that 
$\tilde \tau_1 \chi^0_1 \to \tau \gamma$ can effectively remove neutralinos in 
the early universe. The relative density of $\tilde \tau$ particles to neutralinos 
scales as $e^{-\frac{m_{\tilde \tau} - m_{\chi}}{m_{\chi}}}$, so this scenario tightly 
constrain the $m_{\tilde \tau} - m_{\chi}$ mass difference. Here, the precise 
mass determinations characteristic of LCC1 will not be available: at 0.5~TeV, the ILC 
will observe a single final state, $\tau^+ \tau^- + E_{missing}$, from the two accessible 
SUSY processes~\cite{Khotilovich:2005gb},
$e^+e^- \to \tilde \tau_1 \tilde \tau_1$, $\tilde \tau \to \tau \chi^0_1$ and 
$e^+e^- \to \chi^0_1 \chi^0_2$, $\chi^0_2 \to \chi^0_1 \tilde \tau \to \chi^0_1 \chi^0_1 \tau \tau$. 
The signal topology consists of two $\tau$-jets and missing energy. Background processes, such
as $e^+e^- \to ZZ$ can be suppressed using cuts on event shape variables. The mass splitting 
can be determined by a study of the distribution of the invariant mass of the system made by the 
two $\tau$-jets and the missing energy vector, $M_{j_1 j_2 E_{missing}}$. In this variable, the 
remaining SM background is confined to low values and the shape and upper endpoint of the
$\tilde \tau_1 \tilde \tau_1$ contribution depends on the stau-neutralino mass difference,
$\Delta M = M_{\tilde \tau_1} - M_{\chi^0_1}$. Templates functions can be generated for 
different values of $\Delta M$ and the mass difference is extracted by a $\chi^2$ fit of 
these templates to the ``data''. As the $\Delta M$ value decreases, the energy available to the 
$\tau$ leptons decreases. Since $\tau$ decays involve neutrinos, additional energy is lost from 
detection. When the $\tau \tau$ system becomes soft, the four fermion background process 
$ee \to ee \tau \tau$, the so-called $\gamma \gamma$ background which has cross sections 
at the nb level, makes its detection increasingly difficult. What makes possible to reject these 
$\gamma \gamma$ events is the presence of the two energetic primary electrons at small angle 
w.r.t.\ the beamline~\cite{Bambade:2004tq}. This is a significant challenge for low angle 
calorimetry, since the electron has to be detected in an hostile environment populated by a 
large number of other electrons, of lower energy, arising from pairs created during the bunch 
collision~\cite{Chen:1989ss,Tauchi:1993tm}. A detailed study~\cite{Khotilovich:2005gb}, 
performed for a statistics of 500~fb$^{-1}$, shows that values of $\Delta M$ as small as 5~GeV 
can be measured at the ILC, provided the primary electrons can be vetoed down to 17~mrad. 
In the specific case of the LCC3 point, where the mass splitting, $\Delta M$, is 10.8~GeV, 
an accuracy of 1~GeV can be achieved. Heavier gauginos, as well as the $A^0$ boson, become 
accessible operating the 
ILC at 1~TeV. These data constrain both the mixing angles and $\tan \beta$. As a result the 
neutralino relic density can be estimated with an 18~\% accuracy.
Finally, the LCC4 point, chosen in the $A$ funnel, has the DM density controlled by the 
$\chi \chi \to A$ process. This point is rather instructive in terms of the discovery-driven 
evolution of a possible experimental program at the ILC. The ILC can obtain the neutralino 
and $\tilde \tau$ masses at 0.5~TeV, following the same technique as for LCC3. We would also
expect LHC experiments to have observed the $A^0$ boson, but it is unlikely 
$M_A$ could be determined accurately in $pp$ collisions, since the available observation 
mode is the decay in $\tau$ lepton pairs. At this stage, it would be apparent that the mass
relation between the neutralino mass, accurately measured by the ILC at 0.5~TeV, and the 
$A$ boson mass, from the LHC data, is compatible with $M_A \simeq 2 M_{\chi}$, as required 
for the s-channel annihilation process to be effective. Three more measurements have to be 
performed at the ILC: the $A^0$ mass, $M_A$, and width $\Gamma_A$ and the $\mu$ 
parameter, which is accessible through the mass splitting between heavier neutralinos, 
$\chi^0_3$, $\chi^0_4$ and the lighter $\chi^0_1$, $\chi^0_2$. All these measurements are 
available by operating the ILC at 1~TeV. $M_A$ and $\Gamma_A$ can be determined by studying 
the $A^0$ production in association with a $H^0$ boson, in the reaction 
$e^+e^- \to A^0 H^0 \to b \bar b b \bar b$. 
This process results in spectacular events with four $b$ jets, 
emitted almost symmetrically, due to low energy carried by the heavy Higgs bosons 
(see Figure~\ref{fig12}a). The cross section, for the 
parameters of LCC4 corresponding to $M_A$ = 419~GeV, is just 0.9~fb highlighting the 
need of large luminosity at the highest energy.
Jet flavour tagging and event shape analysis significantly reduces the major multi-jet 
backgrounds, such as $WW$, $ZZ$ and $t \bar t$. The SM $b \bar b b \bar b$ electro-weak 
background has a cross section of $\sim$3~fb, but since it includes $Z^0$ or $h^0$ as
intermediate states it can be efficiently removed by event shape and mass cuts. 
After event selection, the $A^0$ mass and width must be reconstructed from 
the measured di-jet invariant masses. This is achieved 
by pairing jets in the way that minimises the resulting di-jet mass difference, since the 
masses of the $A$ and $H$ bosons are expected to be degenerate within a few GeV, and the 
di-jet masses are computed by imposing constraints on energy and momentum conservation to 
improve the achievable resolution and gain sensitivity to the boson natural width 
(see Figure~\ref{fig12}b). The result is a determination of the $A$ mass to 0.2~\% and of 
its width to $\simeq$15~\% if a sample of 2~ab$^{-1}$ of data can be collected. The full 
set of ILC data provides a neutralino relic density evaluation with 19~\% relative 
accuracy. The full details of how these numbers were obtained can be found in 
Ref.~\cite{Baltz:2006fm} . 

SUSY offers a compelling example for investigating the complementarity 
in the search and discovery of new particles and in the study of their 
properties at the LHC and ILC. The connection to cosmology, through the 
study of dark matter brings precise requirements in terms of accuracy 
and completeness of the anticipated measurements and puts emphasis on 
scenarios at the edges of the parameter phase space. 
The interplay of satellite, ground-based and collider experiments in 
cosmology and particle physics will be unique and it will lead us to 
learn more about the structure of our Galaxy and of the Universe as 
well as of the underlying fundamental laws of the elementary particles. 
This quest will represent an major effort for science in the next 
several decades. The scenarios discussed above highlight the essential 
role of the ILC in this context. It will testing whether the particles 
observed at accelerators are responsible for making up a sizeable fraction 
of the mass of the Universe, through precision spectroscopic measurements.
The data obtained at the ILC will effectively remove most particle physics 
uncertainties and become a solid ground for studying dark matter in our
galaxy through direct and indirect detection experiments~\cite{Feng:2005nz}.

\subsection{Indirect Sensitivity to New Physics at the ILC}\label{sec3.3}

Beyond Supersymmetry there is a wide range of physics scenarios invoking new phenomena 
at, and beyond, the TeV scale. These may explain the origin of electro-weak symmetry 
breaking, if there is no light elementary Higgs boson, stabilise the SM, if SUSY is not 
realised in nature, or embed the SM in a theory of grand unification. The ILC, operating 
at high energy, represents an ideal laboratory for studying this New 
Physics in ways that are complementary to the LHC~\cite{zp1,dominici}. 
Not only it may directly produce some of the new 
particles predicted by these theories, the ILC also retains an indirect sensitivity, 
through precision measurements of virtual corrections to electro-weak observables, 
when the new particle masses exceed the available centre-of-mass energy. 

One of the simplest of such SM extensions consists of the introduction of an 
additional $U(1)$ gauge symmetry, as predicted in some grand unified 
theories~\cite{Hewett:1993st,Rizzo:2006nw}.
The extra $Z'$ boson, associated to the symmetry, naturally mixes with the SM $Z^0$. 
The mixing angle is already strongly constrained, by precision electroweak data, and 
can be of the order of few mrad at most, while direct searches at Tevatron for a new $Z'$ 
boson set a lower limit on its mass around 800~GeV, which may reach 1~TeV by the time the 
LHC will start searching for such a state. 
The search for an extended gauge sector offers an interesting framework 
for studying the ILC sensitivity to scales beyond those directly accessible. It also raises 
the issue of the discrimination between different models, once a signal would be detected. 
The main classes of models with  additional $Z'$ bosons include $E_6$ inspired models and 
left-right models (LR). In the $E_6$ models, the $Z'$ fermion couplings depend on the angle, 
$\theta_6$, defining the embedding of the extra $U(1)$ in the $E_6$ group.
At the ILC, the indirect sensitivity to the mass of the new boson, $M_{Z'}$, can be 
parametrised in terms of the available integrated luminosity, ${\cal{L}}$, and centre-of-mass 
energy, $\sqrt{s}$. A scaling law for large values of $M_{Z'}$ can be obtained by considering 
the effect of the $Z'-\gamma$ interference in the two fermion production cross section 
$\sigma {\mathrm{(}}e^+e^- \to f \bar f {\mathrm{)}}$ ($\sigma_{f \bar f}$ in the following). 
For $s<< M_{Z'}^2$ and assuming the 
uncertainties $\delta \sigma$ to be statistically dominated, we obtain the following 
scaling for the difference between the SM cross section and that in presence of the $Z'$, 
in units of the statistical accuracy:
\begin{equation}
\frac{|\sigma^{SM}_{f \bar f} - \sigma^{SM+Z'}_{f \bar f}|}{\delta \sigma} \propto \frac{1}{M^2_{Z'}}\sqrt{s {\cal{L}}} 
\end{equation}
from which we can derive that the indirect sensitivity to the $Z'$ mass scales with the square
of the centre-of-mass energy and the luminosity as:
\begin{equation}
M_{Z'} \propto (s {\cal{L}})^{1/4}.
\label{resc}
\end{equation}
In a full analysis, the observables sensitive to new physics contribution in 
two-fermion production are the cross section $\sigma_{f \bar f}$, the forward-backward 
asymmetries $A_{FB}^{f \bar f}$ and the left-right asymmetries $A_{LR}^{f \bar f}$. 
The ILC gives us the possibility to study a large number of reactions, 
$e^+_R e^-_L$, $e^+_R e^-_R \to (u \bar u~+~d \bar d),~s \bar s,~c \bar c,~b \bar b,
~t \bar t,~e^+e^-,~\mu^+\mu^-,~\tau^+\tau^-$ with final states of well defined flavour 
and, in several cases, helicity. In order to achieve this, jet flavour tagging is essential 
to separate $b$ quarks from lighter quarks and $c$ quarks from both $b$ and light quarks. 
Jet-charge and vertex-charge reconstruction 
allows then to tell the quark from the antiquark produced in the same 
event~\cite{Ackerstaff:1997ke,Abe:2004hx}. Similarly to LEP and SLC 
analyses, the forward-backward asymmetry can be obtained from a fit to the flow of the 
jet charge $Q^{jet}$, defined as $Q^{jet} = \frac{\sum_i q_i |p_i T|^k}{\sum_i |p_i T|^k}$, 
where $q_i$ is the particle charge, $p_i$ its momentum, $T$ the jet thrust axis and the sum 
is extended to all the particles in a given jet. Another possible technique uses 
the charge of secondary particles to determine the vertex charge and thus the quark charge. 
The application of this technique to the ILC has been studied in some details in relation 
to the optimisation of the Vertex Tracker~\cite{Hillert:2005rp}. At ILC energies, the
$e^+e^- \rightarrow f \bar f$ cross sections are significantly reduced, compared to those 
at LEP and SLC: at 1~TeV the cross section $\sigma(e^+e^- \to b \bar b)$ is only 100~fb, 
so high luminosity is essential and new experimental issues emerge. At 1~TeV, the ILC 
beamstrahlung parameter doubles compared to 0.5~TeV, beam-beam effects becoming important, and 
the primary $e^+e^-$ collision is accompanied by $\gamma \gamma \rightarrow {\mathrm{hadrons}}$ 
interactions~\cite{Chen:1993db}.
Being mostly confined in the forward regions, this background may reduce the 
polar angle acceptance for quark flavour tagging and dilute the jet charge separation using 
jet charge techniques. The statistical accuracy for the determination of 
$\sigma_{f \bar f}$, $A_{FB}^{f \bar f}$ and $A_{LR}^{f \bar f}$ has been 
studied, for $\mu^+\mu^-$ and $b \bar b$, taking the ILC parameters at
$\sqrt{s}$ = 1~TeV. The additional particles from the $\gamma \gamma$ background 
cause a broadening of the $Q^{jet}$ distribution and thus a dilution of the quark charge 
separation. Detailed full simulation and reconstruction is needed to fully understand 
these effects. 
\begin{figure}[h]
\begin{center}
  \minifig{2.1in}{\epsfig{figure=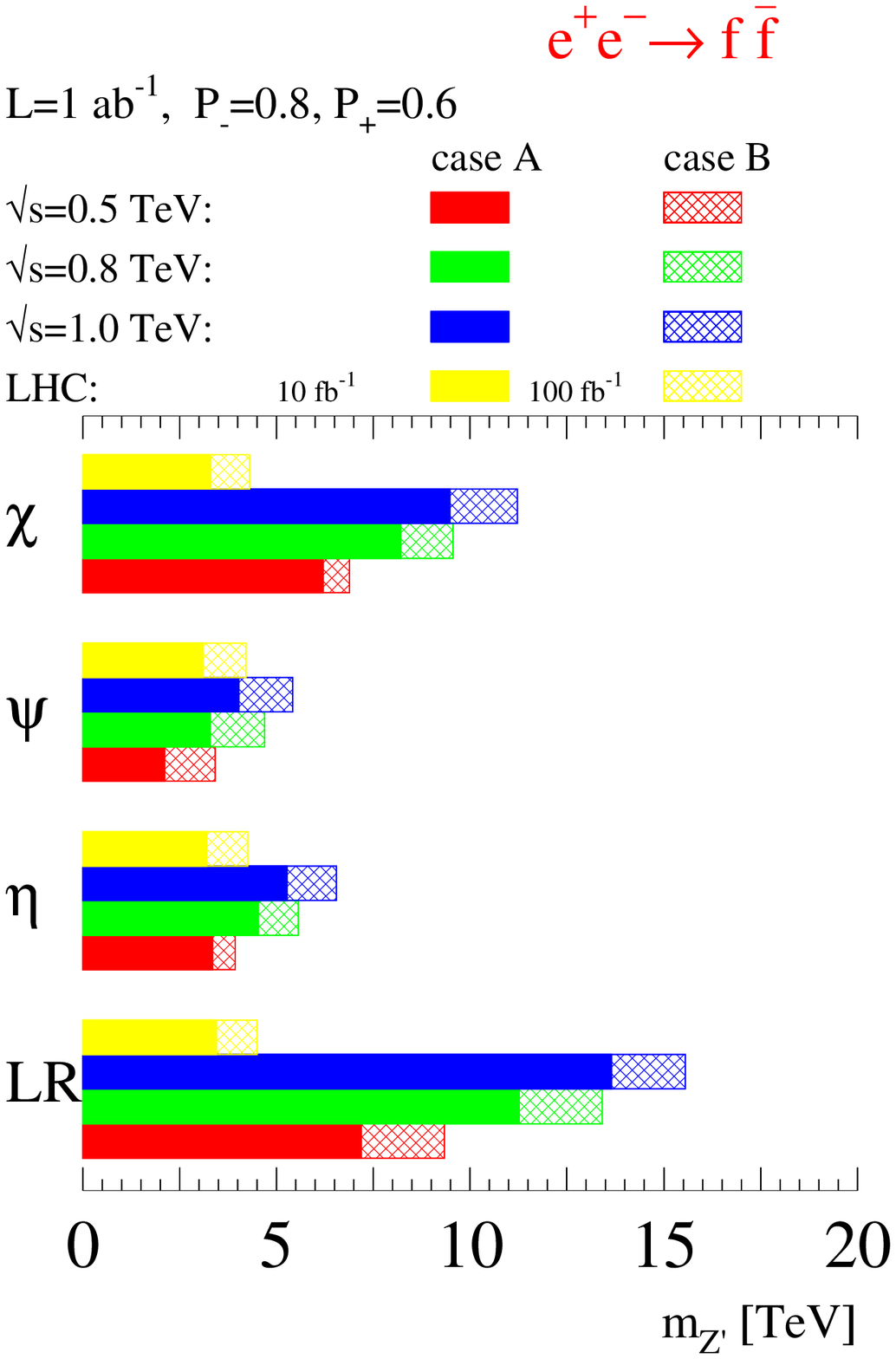,height=6.5cm,clip}}{a}
  \hspace*{1pt}
  \minifig{2.1in}{\epsfig{figure=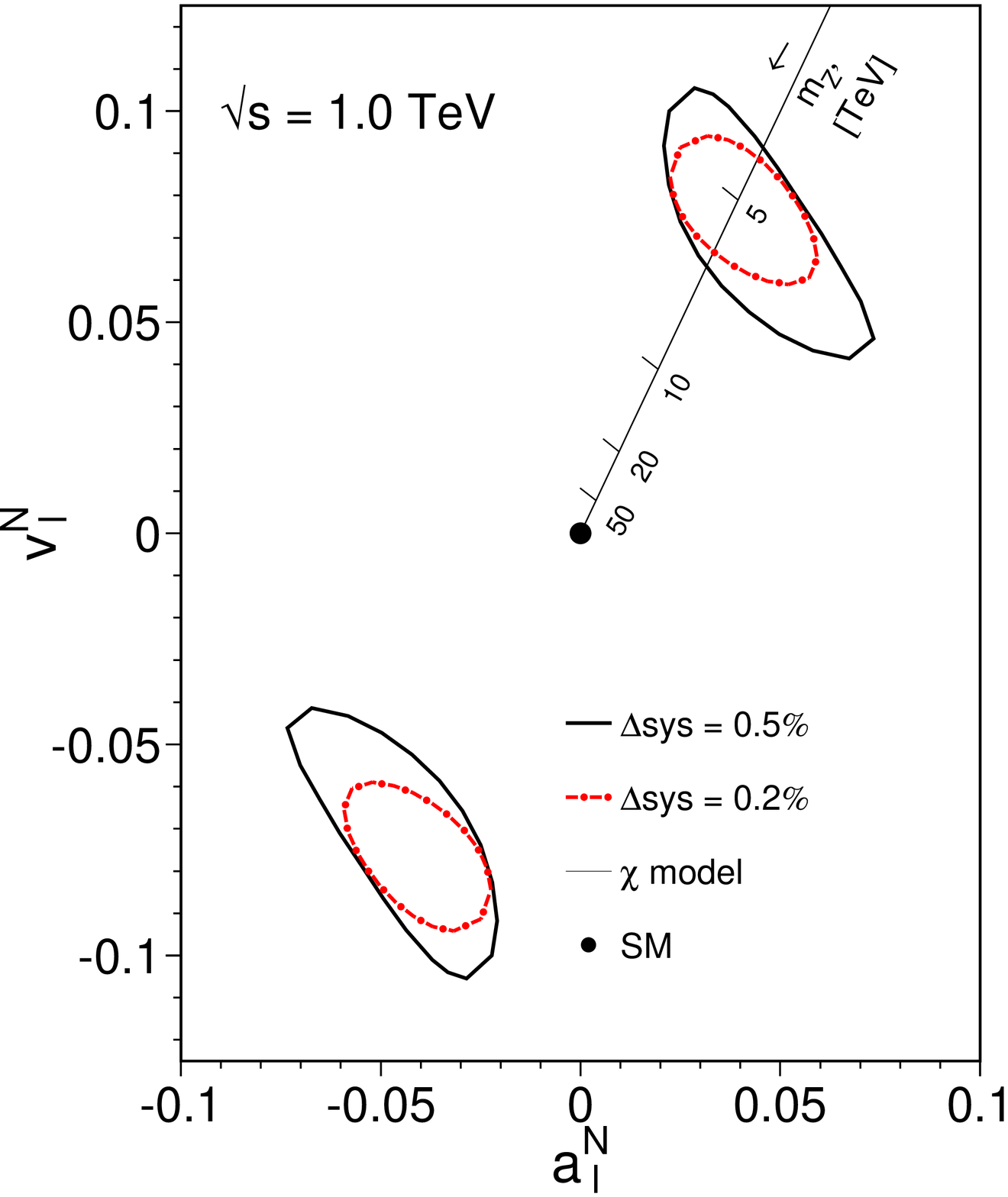,height=6.5cm,clip}}{b}
\caption{Indirect sensitivity to $Z'$ bosons at ILC: (a) mass sensitivity 
to different $Z'$ models for 1~ab$^{-1}$ of data at different centre-of-mass 
energies compared to that of LHC and  (b) accuracy on leptonic couplings for 
a 5~TeV $Z'$ boson (from Ref.~\cite{Aguilar-Saavedra:2001rg})} 
\label{fig14}%
\end{center}
\end{figure}
Despite these backgrounds, the anticipated experimental accuracy in the determination of 
the electro-weak observables in two-fermion processes at 1~TeV is of the order of a few 
percent, confirming the ILC role as the precision machine. Several scenarios of new physics 
have been investigated~\cite{Riemann:1997py,Battaglia:2001fr}.
The analysis of the cross section and asymmetries at 1~TeV would reveal the existence of an 
additional $Z'$ boson up to $\simeq$~6-15~TeV, depending on its couplings.
As a comparison the LHC direct sensitivity extends up to approximately 4-5~TeV. 
The ILC indirect sensitivity also extends to different models on new physics, such as 
5-dimensional extension of the SM with fermions on the boundary for a compactification where 
scales up to about 30~TeV can be explored. 
Finally, fermion compositeness or the exchange of very heavy new particles can be 
described in terms of effective four-fermion contact interactions~\cite{Eichten:1983hw}.
The interaction depends 
on a scale $\Lambda = M_X/g$, where $M_X$ is the mass of the new particle and $g$ the 
coupling. Limits to this scale $\Lambda$ can be set up to $\simeq$~100~TeV, which shows 
that the ILC sensitivity to new phenomena can exceed its centre-of-mass energy by a 
significant factor. In order to maximise this indirect sensitivity to new physics, the 
precision of the SM predictions should match the experimental accuracy. Now, at TeV energies, 
well above the electroweak scale,the ILC will face the effects of large non-perturbative 
corrections. Large logarithms $\propto \alpha^n~log^{2n} (M^2/s)$ arise from the exchange 
of collinear, soft gauge bosons and are known as Sudakov logarithms~\cite{Melles:2001ye}. 
At 1~TeV the logarithmically enhanced $W$ corrections to $\sigma_{b \bar b}$, of the form 
$\alpha~log^2 (M_W^2/s)$ and $\alpha~log (M_W^2/s)$ amount to 19\% and -4\% respectively.  
The effect of these large 
logarithmic corrections has been studied in some 
details~\cite{Ciafaloni:1999ub,Battaglia:2004mw}. 
It will be essential to promote a program of studies to reduce 
these theoretical uncertainties, to fully exploit the ILC potential in these studies.

\subsection{Run Plan Scenario}\label{sec3.4}

One of the points of strength of the ILC is in its remarkable flexibility of running
conditions. Not only the centre-of-mass energy can be changed over approximately an 
order of magnitude, but the beam particle and their polarization state can be varied 
to suit the need of the physics processes under study. At the same time, the ILC program 
is most diversified and data taken at the same centre-of-mass energy may be used for 
very different analyses, such as precise top mass determination, Higgs boson studies 
and reconstruction of SUSY decays. This has raised concerns whether the claimed 
ILC accuracies can be all achieved with a finite amount of data. A dedicated 
study was performed in 2001, under the guidance of Paul Grannis, taking two physics 
scenarios with Supersymmetry realised 
at relatively low mass, one being the LCC1 benchmark point, rich in pair-produced 
particles and requiring detailed threshold scans~\cite{Battaglia:2002ey}. The study
assumes a realistic profile for the delivered luminosity, which increases from 
10~fb$^{-1}$ in the first year to 200~fb$^{-1}$ in the fifth year and 250~fb$^{-1}$
afterward, for a total integrated equivalent luminosity  $\int {\cal{L}}$ = 1~ab$^{-1}$. 
The proposed run plan starts at the assumed maximum energy of 0.5~TeV 
for a first determination of the sparticle masses through the end-point study and 
then scans the relevant thresholds, including $t \bar t$ in short runs with tuned
polarization states. A summary is given in Table~\ref{tab:runplan}.
\begin{table}[h]
\tbl{ILC Run plan scenario for LCC1.}
{\begin{tabular}{@{}ccccl@{}} \toprule
Beams & $\sqrt{s}$ & Pol. & $\int {\cal{L}}$ & Comments \\ 
      & (TeV)      &      & (fb$^{-1}$)      &          \\ \colrule
$e^+e^-$ & 0.500   & L/R   &  335 & Sit at max.\ energy for sparticle endpoint measurements \\
$e^+e^-$ & 0.270   & L/R   &  100 & Scan $\chi^0_1 \chi^0_2$ (R pol.) and 
$\tilde \tau_1 \tilde \tau_1$ (L pol.) \\
$e^+e^-$ & 0.285   & R     &  ~50 & Scan $\tilde \mu_R \tilde \mu_R$ \\
$e^+e^-$ & 0.350   & L/R   &  ~40 & Scan $t \bar t$, $\tilde e_R \tilde e_L$ (L \& R pol.), 
$\chi^+_1 \chi^-_1$ (L pol.)\\
$e^+e^-$ & 0.410   & L/R     &  100 & Scan $\tilde \tau_2 \tilde \tau_2$ \\
$e^-e^-$ & 0.285   & RR     &  ~10 & Scan for $\tilde e_R$ mass \\
\botrule
\end{tabular}}
\label{tab:runplan}
\end{table}
This plan devotes approximately two third of the total luminosity at, or near, the maximum 
energy, so the program will be sensitive to unexpected new phenomena at high energy, 
while providing accurate measurements of masses through dedicated scans.

\section{Sensors and Detectors for the ILC}\label{sec4}

The development of the ILC accelerator components and the definition of its 
physics case has been paralleled by a continuing effort in detector design 
and sensor R\&D.
This effort is motivated by the need to design and construct detectors which 
match the ILC promise to provide extremely accurate measurements over a broad 
range of collision energies and event topologies. It is important to stress 
that, despite more than a decade of detector R\&D for the LHC experiments, much 
still needs to be done to obtain sensors matching the ILC requirements. 
While the focus of the LHC-motivated R\&D has been on sensor radiation hardness
and high trigger rate, the ILC, with its more benign background conditions 
and lower interaction cross sections, admits sensors of new technology which, 
in turn, have better granularity, smaller thickness and much improved resolution.
Sensor R\&D and detector design are being carried out world-wide and are starting
deploying prototype detector modules on test beamlines.     
 
\subsection{Detector Concepts}\label{sec4.1}

The conceptual design effort for an optimal detector for the ILC interaction 
region has probed a wide spectrum of options which span from a spherical 
detector structure to improved versions of more orthodox barrel-shaped detectors.
These studies have been influenced by the experience with SLD at
the SLC, ALEPH, DELPHI and OPAL at LEP, but also with ATLAS and CMS at the LHC. 
The emphasis on accurate reconstruction of the 
particle flow in hadronic events and thus of the energy of partons is 
common to all designs. The main tracker technology drives the detector designs 
presently being studied. Four detector concepts have emerged, named
GLD, LDC, SiD and 4$^{th}$ Concept~\cite{concepts}. 
\begin{figure}[h]%
\begin{center}
\centerline{\epsfig{figure=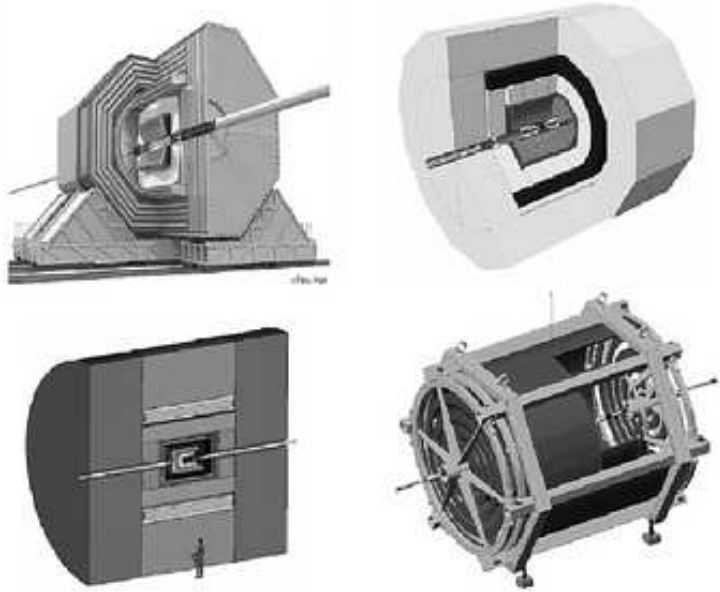,width=10.0cm}}
\caption{View of the four ILC detector concepts presently being studied: GLD (upper left),
LDC (upper right), SiD (lower left) and 4$^{th}$ Concept (lower right).}
\label{fig15}%
\end{center}
\end{figure}
A large volume, 3D continuous tracking volume 
in a Time Projection Chamber is the centerpiece of the GLD, the LDC and the 
so-called 4$^{th}$ Concept designs. 
The TPC is followed by an highly segmented electro-magnetic calorimeter for 
which these three concepts are contemplating different technologies  
A discrete tracker made of layers of high precision Silicon microstrip 
detectors, and a larger solenoidal field, which allows to reduce the radius, 
and thus the size, of the calorimeter is being studied in the context of the 
SiD design. Dedicated detector design studies are being carried out 
internationally~\cite{Behnke:2005re,Abe:2006bz} to optimise, 
through physics benchmarks~\cite{Battaglia:2006bv}, 
the integrated detector concepts. Such design activities provide a bridge 
from physics studies to the assessment of priorities in detector R\&D and are 
evolving towards the completion of engineered design reports at the end of this 
decade, synchronously with that foreseen for the ILC accelerator.

\subsection{Vertexing and Tracking}\label{sec4.2}

The vertex and main tracker detectors must provide jet flavour identification 
and track momentum determination with the accuracy which makes 
the ILC such a unique facility for particle physics. 
The resolution in extrapolating charged particle trajectories to their 
production point, the so-called impact parameter, is dictated by the need 
to distinguish Higgs boson decays to $c \bar c$ from those to $b \bar b$ 
pairs, but also $\tau^+ \tau^-$ and gluon pairs, as discussed in 
section~\ref{sec3.1}. In addition, vertex charge measurements put 
emphasis on precise extrapolation of particle tracks down to very low 
momenta. Tagging of events with multiple $b$ jets, such as 
$e^+e^- \to H^0A^0 \to b \bar b b \bar b$, discussed in section~\ref{sec3.2}, 
underscores the need of high tagging efficiency, $\epsilon_b$, since the 
overall efficiency scales as $\epsilon_b^N$, where $N$ is the number of 
jets to be tagged. This is best achieved by analysing the secondary vertex 
structures in hadronic jets. A $B$ meson, from a Higgs boson produced at 
0.5~TeV, has an average energy of $x_B \sqrt{s}/4 \simeq$ 100~GeV, where 
$x_B \simeq 0.7$ represents the average $b$ fragmentation function, 
or a $\gamma$ value of $\simeq$~70. 
Since $c \tau \simeq$~500~$\mu$m, the average decay 
distance $\beta \gamma c \tau$ is 3.5~mm and the average impact parameter, 
$\beta \gamma c \tau \sin \theta$, is 0.5~mm. In comparison, a $D$
meson from a $H \to c \bar c$ decay has a decay length of 1.3~mm. 
More importantly, the average charged decay multiplicity for a $B$ 
meson is 5.1, while for a $D$ meson is 2.7. Turning these numbers 
into performance requirements sets the target accuracy for the asymptotic 
term $a$ and the multiple scattering term $b$ defining the track extrapolation 
resolution in the formula
\begin{equation}
\sigma_{\mathrm{extrapolation}} = {\mathrm{a}}~\oplus \frac{{\mathrm{b}}}{p_t}
\end{equation}
The ILC target values are compared to those achieved by the DELPHI experiment 
at LEP, those expected for ATLAS at the LHC and the best performance ever 
achieved at a collider experiment, that of SLD, in Table~\ref{tab:ipres}.
\begin{table}[h]
\tbl{Values for the asymptotic term $a$ and multiple scattering term $b$ 
defining the track extrapolation resolution required for the ILC compared 
to those obtained by other collider experiments.}
{\begin{tabular}{@{}lcc@{}} \toprule
Experiment & $a$ ($\mu$m) & $b$ ($\mu$m/GeV) \\ \colrule
ILC & 5 & 10 \\
DELPHI & 28 & 65 \\
ATLAS &  15 & 75 \\
SLD & ~8 & 33 \\
\botrule
\end{tabular}}
\label{tab:ipres}
\end{table}
This comparison shows that the improvements required for ILC on state-of-the-art 
technology is a factor 2-5 on asymptotic resolution and another factor 3-7 on the 
multiple scattering term.

At the ILC, particle tracks in highly collimated jets contribute a local track density 
on the innermost layer of 0.2-1.0~hits mm$^{-2}$ at 0.5~TeV, to reach 
0.4-1.5~hits mm$^{-2}$ at 1.0~TeV. Machine-induced backgrounds, mostly pairs, 
add about 3-4~hits mm$^{-2}$, assuming that the detector integrates 80 consecutive 
bunch crossings in a train. These values are comparable to, 
or even exceed, those expected on the innermost layer of the LHC detectors: 
0.03 hits~mm$^{-2}$ for proton collisions in ATLAS and 0.9 hits~mm$^{-2}$ for 
heavy ion collisions in ALICE. Occupancy and point resolution set the pixel 
size to 20x20~$\mu$m$^2$ or less. The impact parameter accuracy  
sets the layer material budget to $\le 0.15\%~X_0$/layer. This motivates 
the development of thin monolithic pixel sensors. Charge coupled devices (CCD) 
have been a prototype architecture after the success of the SLD 
VXD3~\cite{Abe:1999ky}. However, to match the ILC requirements in terms of 
radiation hardness and read-out speed significant R\&D is needed. New technologies, 
such as CMOS active pixels~\cite{Turchetta:2001dy}, 
SOI~\cite{Marczewski:2005vy} and DEPFET~\cite{Richter:2003dn} sensors, 
are emerging as promising, competitive alternatives, supported by an intensive 
sensor R\&D effort promoted for the ILC~\cite{Battaglia:2003kn}. 

The process $e^+e^- \to H^0Z^0$, $H^0 \to X$, 
$Z^0 \to \ell^+ \ell^-$ gives access to Higgs production, irrespective of 
the Higgs decay properties. Lepton momenta must be measured very accurately 
for the recoil mass resolution to be limited by the irreducible smearing due 
to beamstrahlung. Since the centre-of-mass energy $\sqrt{s} = E_{H} + E_{Z}$ 
is known and the total momentum $p_{H} + p_{Z} = 0$, the Higgs mass, $M_H$ 
can be written as:
\begin{equation}
M_H^2 = E_H^2 - p_H^2 = (\sqrt{s} - E_Z)^2 - p_Z^2 = s + E_Z^2 - 2\sqrt{s} E_Z 
- p_Z^2 = s-2\sqrt{s} E_Z + M_Z^2 
\end{equation}
In the decay $Z^0 \to \mu^+ \mu^-$, $E_Z = E_{\mu^+} + E_{\mu^-}$ so that the 
resolution on $M_H$ depends on that on the muon momentum. In quantitative terms 
the resolution required is
\begin{equation}
\delta p / p^2 < 2 \times 10^{-5}
\end{equation}
A comparison with the performance of trackers at LEP and LHC is given in 
Table~\ref{tab:pres}.
\begin{table}[h]
\tbl{Values for the momentum resolution $\delta p / p^2$ for the main tracker and the 
full tracking system at ILC, LEP and LHC. These values do not include the vertex 
constraint.}
{\begin{tabular}{@{}lcc@{}} \toprule
Experiment & Main Tracker & Full Tracker \\
           & Only         &              \\
ILC & 1.5 $\times$ 10$^{-4}$ & 5 $\times$ 10$^{-5}$ \\
ALEPH & 1.2 $\times$ 10$^{-3}$ & 5 $\times$ 10$^{-4}$ \\
ATLAS & -- & 2 $\times$ 10$^{-4}$ \\
\botrule
\end{tabular}}
\label{tab:pres}
\end{table}
The ability to tag Higgs bosons, independent on their decay mode is central to 
the ILC program in Higgs physics. A degraded momentum resolution would correspond 
to larger background, mostly from $e^+e^- \to ZZ^*$, being accepted in the Higgs 
signal sample. This degrades the accuracy on the determination of the Higgs 
couplings both in terms of statistical and systematic uncertainties. 
The particle momentum is measured through its bending radius $R$ in the solenoidal 
magnetic field, $B$. The error on the curvature, $k=1/R$, for a particle track 
of high momentum, measured at $N$ equidistant points with an accuracy, $\sigma$, 
over a length L, applying the constraint that it does originate at the primary 
vertex (as for the leptons from the $Z^0$ in the Higgstrahlung reaction) is given 
by~\cite{pdg}:
\begin{equation}
\delta k = \frac{\sigma}{L^2}\sqrt{\frac{320}{N+4}}
\end{equation}
This shows that the same momentum resolution can be achieved either by a large 
number of measurements, each of moderate accuracy, as in the case of a 
continuous gaseous tracker, or by a small number of points measured with high 
accuracy, as in the case of a discrete Si tracker.
Continuous tracking capability over a large area, with timing 
information and specific ionization measurement, and its robust 
performance make the Time Projection Chamber an attractive option for 
precision tracking at the ILC. The introduction of Micro Pattern Gaseous 
Detectors~\cite{Giomataris:1995fq,Sauli:1997qp} 
(MPGD) offers significant improvements in terms of reduced 
$E \times B$, larger gains, ion suppression and faster, narrower signals 
providing better space resolution. Improving on the space resolution 
requires an optimal sampling of the collected charge, while the high 
solenoidal magnetic field reduces the diffusion effects. Several paths are 
presently being explored with small size prototypes operated on beamlines 
and in large magnetic fields~\cite{Kappler:2004cg,Colas:2004ks}.

A multi-layered Si strip detector tracker in an high $B$ field may offer a 
competitive $\delta p/p^2$ resolution with reduced material budget and afford 
a smaller radius ECAL, thus reducing the overall detector cost. 
This is the main rationale promoting the development of an all-Si concept for 
the main tracker, which follows the spirit of the design of the CMS detector 
at LHC. Dedicated conceptual design and module R\&D is being carried out as a 
world-wide program~\cite{Kroseberg:2005ue}.
There is also considerable R\&D required for the engineering 
of detector ladders, addressing such issues as mechanical stability and 
integration of cooling and electrical services. These modules may also be 
considered as supplemental tracking devices in a TPC-based design to provide 
extra space points, with high resolution, and in end-cap tracking planes. 
Assessing the required detector performance involves realistic simulation and 
reconstruction code accounting for inefficiencies, noise, overlaps and 
backgrounds. 

\subsection{Calorimetry}\label{sec4.3}

The ILC physics program requires precise measurements of multi-jet hadronic 
events, in particular di-jet invariant masses to identify $W$, $Z$ and Higgs 
bosons, through their hadronic decays. An especially demanding reaction is 
$e^+e^- \to Z^0 H^0 H^0$, which provides access to the triple Higgs coupling
as discussed in section~\ref{sec3.1}. The large background from 
$e^+e^- \to Z^0 Z^0 Z^0$ can be reduced only by an efficient $H^0$/$Z^0$ 
separation, based on their masses. This impacts the parton energy resolution 
through the measurement of hadronic jets. Detailed simulation~\cite{Castanier:2001sf} 
shows that a jet energy resolution 
$\frac{\sigma_{E_{jet}}}{E_{jet}} \simeq \frac{0.30}{\sqrt{E}}$
is required, in order to achieve an interesting resolution on the $g_{HHH}$ coupling.
The analysis of other processes, such as $e^+e^- \to W^+W^- \nu \bar \nu$ and Higgs
hadronic decays, leads to similar conclusions~\cite{Brient:2002gh}.
In the case of the determination of $H^0 \to W^+W^-$ branching fractions, the statistical 
accuracy degrades by 22~\% when changing the jet energy resolution from 
$\frac{0.30}{\sqrt{E}}$ to $\frac{0.60}{\sqrt{E}}$.
\begin{figure}[h]%
\begin{center}
  \parbox{2.1in}{\epsfig{figure=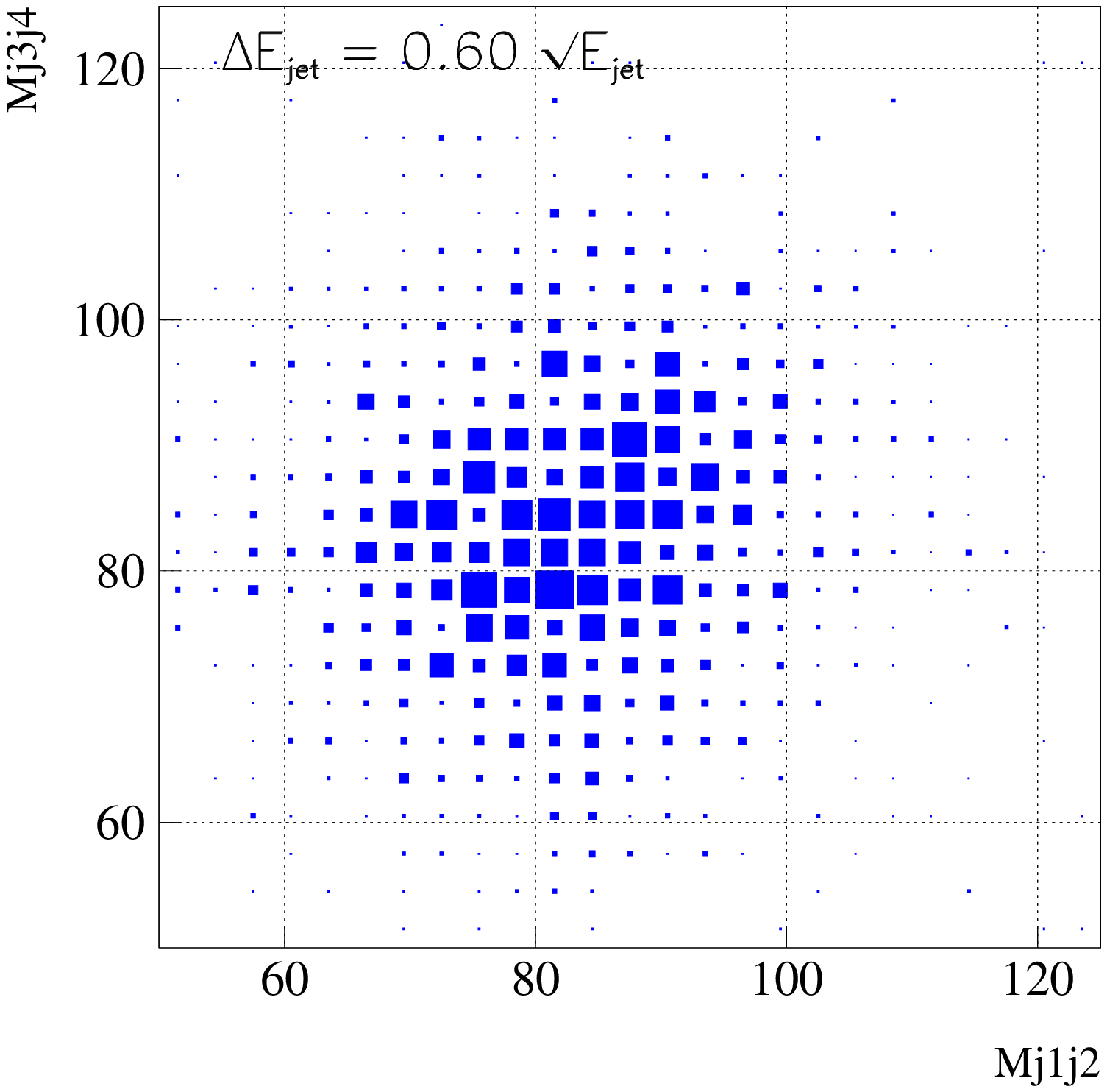,width=5.5cm,height=6.5cm}
  \figsubcap{a}}
  \hspace*{1pt}
  \parbox{2.1in}{\epsfig{figure=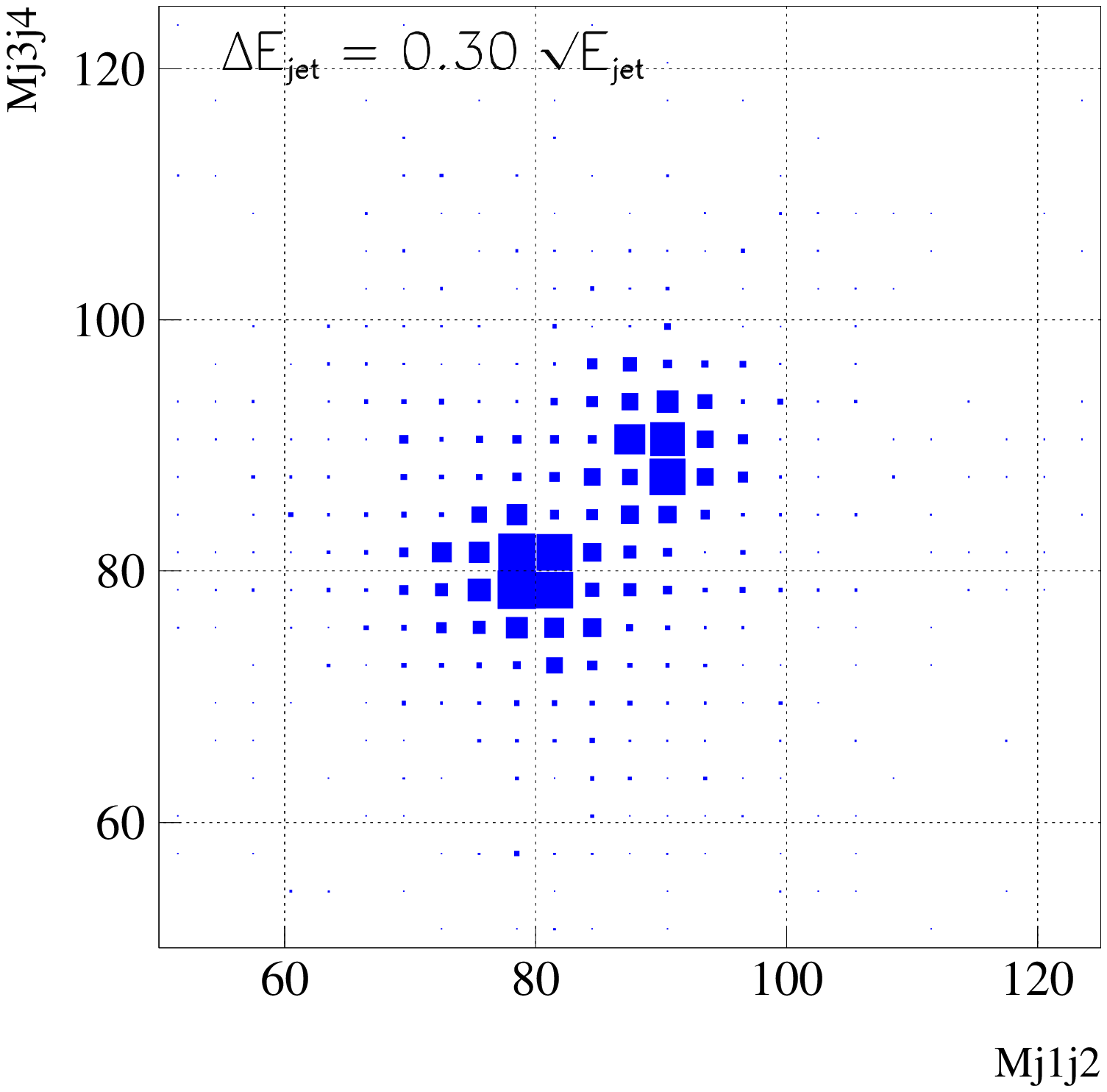,width=5.5cm,height=6.5cm}
  \figsubcap{b}}
  \caption{$W^{\pm}$ and $Z^0$ gauge boson pair production separation at the ILC: 
invariant mass of the first di-jet pair vs. that of the second for a sample of $WW$ 
and $ZZ$ for two different assumptions on the jet energy resolution
(a) $\frac{0.30}{\sqrt{E}}$ and (b) $\frac{0.60}{\sqrt{E}}$
(from Ref~\cite{Behnke:2001qq}).}%
  \label{fig17}%
\end{center}
\end{figure}
Such performance is unprecedented and requires the development of an advanced calorimeter 
design as well as new reconstruction strategies. The most promising approach is based on 
the {\it particle flow algorithm} (PFA). 
The energy of each particle in an hadronic jet is determined based on the information 
of the detector which can measure it to the best accuracy. In the case of charged particles, 
this is achieved by measuring the particle bending in the solenoidal field with the main 
tracker. Electromagnetic neutrals ($\gamma$ and $\pi^0$) are measured in the electromagnetic 
calorimeter and hadronic neutrals ($K^0_L$, $n$) in the hadronic calorimeter. The 
jet energy is then obtained by summing these energies:
\begin{equation}
E_{jet} = E_{charged} + E_{em~neutral} + E_{had~neutral}
\end{equation}
each being measured in a specialised detector. The resolution is given by:
\begin{equation}
\sigma^2_{E_{jet}} = \sigma^2_{charged} + \sigma^2_{em~neutral} + \sigma^2_{had~neutral} + 
\sigma^2_{confusion}.
\end{equation}
Assuming the anticipated momentum resolution, $\sigma_{E} \simeq 0.11/\sqrt{E}$ for the 
e.m.\ calorimeter,  $\sigma_E \simeq 0.40/\sqrt{E}$ for the hadronic calorimeter and the 
fractions of charged, e.m. neutral and hadronic neutral energy in an hadronic jet we get:
\begin{equation}
\sigma^2_{charged} \simeq (0.02 {\mathrm{GeV}})^2 \frac{1}{10} 
\sum \frac{E^4_{charged}}{(10 {\mathrm{GeV}})^4}
\end{equation}
\begin{equation}
\sigma^2_{em~neutral} \simeq (0.6 {\mathrm{GeV}})^2 \frac{E_{jet}}{100 {\mathrm{GeV}}}
\end{equation}
\begin{equation}
\sigma^2_{had~neutral} \simeq (1.3 {\mathrm{GeV}})^2 \frac{E_{jet}}{100 {\mathrm{GeV}}}
\end{equation}
In case of perfect energy-particle association this would correspond to a jet resolution
$\simeq 0.14/\sqrt{E}$. But a major source of resolution loss turns out to be the 
confusion term, $\sigma_{confusion}$, which originates from inefficiencies, double-counting 
and fakes, which need to be minimised by an efficient pattern recognition. This strategy was
pioneered by the ALEPH experiment at LEP, where a resolution $\simeq 0.60/\sqrt{E}$ was 
obtained, starting from the stochastic resolutions of $\sigma_{E} \simeq 0.18/\sqrt{E}$ for 
the e.m.\ calorimeter,  and $\sigma_E \simeq 0.85/\sqrt{E}$ for the hadronic 
calorimeter~\cite{Buskulic:1994wz}.
At hadron colliders, the possible improvement from using tracking information together 
with calorimetric measurements is limited, due to underlying events and the shower 
core size. On the contrary, at the ILC these limitations can be overcome, by developing 
an imaging calorimeter, where spatial resolution becomes as important as energy resolution. 
\begin{figure}[h]%
\begin{center}
\epsfig{figure=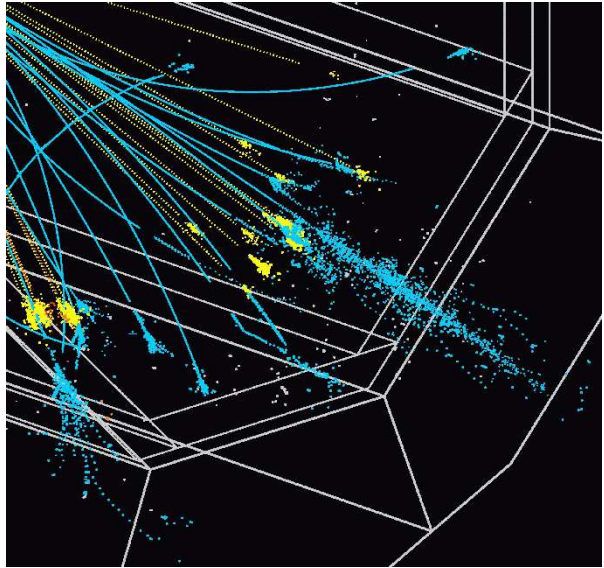,height=7.5cm}
\caption{Visualisation of the imaging calorimeter for the ILC: simulated response of a SiW 
calorimeter to a jet from $e^+e^- \to W^+W^- \to {\mathrm jets}$ at $\sqrt{s}$=0.8~TeV.}%
  \label{fig18}%
\end{center}
\end{figure}
The minimisation of the confusion rate can then be obtained by choosing a large solenoidal 
field, $B$, and calorimeter radius, $R$, to increase the separation between charged 
and neutral particles in dense jets, a small Moliere radius, $R_M$, for the e.m.\ calorimeter, 
to reduce the transverse shower spread and small cells, $R_{pixel}$, with large longitudinal 
segmentation. The distance between a neutral and a charged particle, of transverse momentum 
$p_t$, at the entrance of the e.m.\ calorimeter located at a radius $R$ is given by 
$0.15 B R^2 / p_t$, where $B$ is the solenoidal magnetic field. A useful figure of merit of 
the detector in terms of the particle flow reconstruction capability is then offered by:
\begin{equation}
\frac{B R^2}{R_M^2 R_{pixel}^2}
\end{equation}
which is a measure of the particle separation capability.
The value of $BR^2$ is limited to about 60~Tm$^2$ by the mechanical stability. An 
optimal material in terms of Moliere radius is Tungsten, with $R_M$ = 9~mm. 
In four-jet events at $\sqrt{s}$=0.8~TeV, there are on average 28~GeV per di-jet 
carried by photons, which are deposited within 2.5~cm from a charged particle at the 
e.m.\ calorimeter radius. 
With pixel cells of order of $1 \times 1$~cm$^2$ to ensure sufficient transverse 
segmentation and 30 to 40 layers in depth, the e.m.\ calorimeter would consists 
of up to 30~M channels and 3000~m$^2$ of active Si. Due to the large 
amount of channels and the wish to use an absorber with the smallest possible 
Moliere radius, the e.m.\ calorimeter is the main cost-driver of the ILC detector
and its optimisation in terms of performance and cost requires a significant 
R\&D effort. A Silicon-Tungsten calorimeter (SiW) was first proposed in the framework 
of the TESLA study~\cite{Videau:2000es,Behnke:2001qq} and it is currently being pursued 
by large R\&D collaborations in both Europe and the US. 
Alternative technologies are also being studied by the GLD and the 4$^{th}$ 
Concept. This R\&D program involves design, prototyping and tests with high energy 
particle beams and it is being carried out 
world-wide~\cite{Strom:2005id,Mavromanolakis:2005yh,Strom:2005xt}, 
supported by efforts on detailed simulation and reconstruction.

\section{Epilogue}

The ILC promises to complement and expand the probe into the TeV scale
beyond the LHC capabilities, matching and improving its energy reach 
while adding precision. Its physics program will address many of the 
fundamental questions of today's physics from the origin of mass, to the 
nature of Dark Matter. After more than two decades of intense R\&D carried 
out world-wide, the $e^+e^-$ linear collider, with centre-of-mass energies 
up to 1~TeV, has become technically feasible and a costed reference design 
is now available. Detectors matching the precision requirements of its 
anticipated physics program are being developed in an intense R\&D 
effort carried out world-wide. Now, theoretical predictions matching 
the anticipated experimental accuracies are crucially needed, as well as 
further clues on what physics scenarios could be unveiled by signals that
the LHC may soon be observing. These will contribute to further define the 
physics landscape for the ILC. A TeV-scale electron-positron linear collider 
is an essential component of the research program that will provide in the 
next decades new insights into the structure of space, time, matter and energy. 
Thanks to the efforts of many groups from laboratories and universities 
around the world, the technology for achieving this goal is now in hand, 
and the prospects for the ILC success are extraordinarily bright.	

\section*{Acknowledgments}
I am grateful to the TASI organisers, in particular to Sally Dawson and Rabindra 
N.\ Mohapatra, for their invitation and the excellent organization. I am indebted to 
many colleagues who have shared with me both the excitement of the ILC physics studies 
and detector R\&D, over many years, as well as many of the results included in this 
article. I would like to mention here Ugo Amaldi, Timothy Barklow, Genevieve Belanger, 
Devis Contarato, Stefania De~Curtis, Jean-Pierre Delahaye, Albert De Roeck, Klaus Desch, 
Daniele Dominici, John Ellis, JoAnne Hewett, Konstantin Matchev, Michael Peskin and 
Tom Rizzo. I am also grateful to Barry Barish, JoAnne Hewett, Mark Oreglia and Michael 
Peskin for reviewing the manuscript and their suggestions.

This work was supported in part by the Director, Office of Science, of the
U.S. Department of Energy under Contract No.DE-AC02-05CH11231.

\printindex

\begin{thebibliography}{99}

\bibitem{Tigner:1965}
  M.~Tigner, \emph{Nuovo Cim.}, {\bf 37} 1228 (1965).

\bibitem{Amaldi:1976}
  U.~Amaldi, \emph{Phys. Lett.} {\bf B61} 313  (1976).

\bibitem{Richter:1979cq}
  B.~Richter,
  IEEE Trans.\ Nucl.\ Sci.\  {\bf 26}, 4261 (1979).

\bibitem{Schnell:1986ig}
  W.~Schnell, {\it A Two Stage Rf Linear Collider Using 
  A Superconducting Drive Linac}, CERN-LEP-RF/86-06 (1976).

\bibitem{Ahn:1988vj}
  C.~r.~Ahn {\it et al.},
  {\it Opportunities and Requirements for Experimentation at a Very High-Eenergy $e^+e^-$ Collider}, 
  SLAC-0329 (1988).

\bibitem{trc}
  G.~Loew (editor), \emph{International Linear Collider Technical Review Committee: 
  Second Report}, SLAC-R-606 (2003).

\bibitem{itrp}
{\begin{verbatim}http://www.ligo.caltech.edu/~skammer/ITRP_Home.html \end{verbatim}}

\bibitem{gde}
{\tt http://www.linearcollider.org/}

\bibitem{rdr}
{\tt http://media.linearcollider.org/rdr\_draft\_v1.pdf}

\bibitem{oecd}
Report of the OECD Consultative Group on High-Energy Physics, June 2002
({\tt http://www.oecd.org/dataoecd/2/32/1944269.pdf})

\bibitem{epp2010}
{\tt http://www7.nationalacademies.org/bpa/EPP2010.html}

\bibitem{Maiani:2001wi}
  L.~Maiani,
  prepared for the {\it 9th International Symposium on Neutrino Telescopes}, 
  Venice, Italy, 6-9 March 2001.

\bibitem{Treille:2002iu}
  D.~Treille,
  \emph{Nucl.\ Phys.\ Proc.\ Suppl.}  {\bf 109B}, 1 (2002).

\bibitem{Brinkmann:2001qn}
  R.~Brinkmann, K.~Flottmann, J.~Rossbach, P.~Schmueser, N.~Walker and H.~Weise
(editors), \emph{TESLA: The superconducting electron positron linear collider 
 with an integrated X-ray laser laboratory. Technical design report.}, 
 DESY-01-011B (2001).

\bibitem{Assmann:2000hg}
  R.~W.~Assmann {\it et al.},
  {\it A 3-TeV $e^+e^-$ linear collider based on CLIC technology},
  CERN-2000-008 (2000).

\bibitem{wuensch}
  W.~Wuensch, {\it Progress in Understanding the High-Gradient Limitations of 
  Accelerating Structures}, CLIC-Note-706 (2007).

\bibitem{Battaglia:2004mw}
  M.~Battaglia, A.~De Roeck, J.~Ellis and D.~Schulte (editors),
  {\it Physics at the CLIC multi-TeV linear collider: Report of the CLIC  Physics
  Working Group}, CERN-2004-005 (2004) and arXiv:hep-ph/0412251.

\bibitem{Leemans:2006}
  W.P.~Leemans {\it et al.}, \emph{Nature Physics} {\bf 2} 696 (2006).

\bibitem{Richter:1981uu}
  B.~Richter, SLAC-PUB-2854 (1981)

\bibitem{Amaldi:1987xt}
  U.~Amaldi, {\it Summary talk given at Workshop on Physics at Future 
  Accelerators, La Thuile, Italy, Jan 7-13, 1987}, CERN-EP/87-95 (1987).

\bibitem{Noble:1986yz}
  R.~J.~Noble,
  Nucl.\ Instrum.\ Meth.\  A {\bf 256}, 427 (1987).

\bibitem{Murayama:1996ec}
  H.~Murayama and M.~E.~Peskin,
  Ann.\ Rev.\ Nucl.\ Part.\ Sci.\  {\bf 46}, 533 (1996)
  [arXiv:hep-ex/9606003].

\bibitem{Aguilar-Saavedra:2001rg}
  J.~A.~Aguilar-Saavedra {\it et al.}  [ECFA/DESY LC Physics Working Group],
  \emph{TESLA Technical Design Report Part III: Physics at an e+e- Linear
  Collider}, 
  DESY-2001-011C (2001) and arXiv:hep-ph/0106315.

\bibitem{Abe:2001wn}
  T.~Abe {\it et al.}  [American Linear Collider Working Group],
  {\it Linear collider physics resource book for Snowmass 2001}, 
  SLAC-R-570 (2001).

\bibitem{Abe:2001gc}
  K.~Abe {\it et al.}  [ACFA Linear Collider Working Group],
  {\it Particle physics experiments at JLC}, 
  KEK-REPORT-2001-11 (2001) and arXiv:hep-ph/0109166.

\bibitem{Dawson:2004xz}
  S.~Dawson and M.~Oreglia,
  Ann.\ Rev.\ Nucl.\ Part.\ Sci.\  {\bf 54}, 269 (2004)
  [arXiv:hep-ph/0403015].

\bibitem{Higgs} 
  P.W.~Higgs, \emph{Phys. Rev. Lett.} {\bf 12} 132  (1964);
  {\it idem}, \emph{Phys. Rev.}  {\bf 145} 1156 (1966); F.~Englert and R.~Brout, 
  {\it Phys. Rev. Lett.} {\bf 13} 321 (1964); G.S.~Guralnik, C.R.~Hagen and 
  T.W.~Kibble, \emph{Phys. Rev. Lett.} {\bf 13} 585 (1964).

\bibitem{triv} 
  A.~Hasenfratz {\it et al.}, \emph{Phys. Lett.} {\bf B199} 531
  (1987); M.~L\"uscher and P.~Weisz, \emph{Phys. Lett.} {\bf B212} 472 (1988);
  M.~G\"ockeler {\it et al.}, \emph{Nucl. Phys.} {\bf B404} 517 (1993).

\bibitem{Barate:2003sz}
  R.~Barate {\it et al.}  [LEP Working Group for Higgs boson searches],
 \emph{Phys.\ Lett.\ B} {\bf 565}, 61 (2003) [arXiv:hep-ex/0306033].

\bibitem{ewwg:2005di}
  LEP Electroweak Working Group, Report CERN-PH-EP-2006 (2006),
  arXiv:hep-ex/0612034 and subsequent updates available at 
  {\tt http://lepewwg.web.cern.ch/LEPEWWG/}.

\bibitem{Heinemeyer:2005gs}
  S.~Heinemeyer {\it et al.},
  arXiv:hep-ph/0511332.

\bibitem{hmass1}
  P.~Garcia-Abia, W.~Lohmann and A.~Raspereza,
  Note LC-PHSM-2000-062 (2000).

\bibitem{Miller:2001bi}
  D.~J.~Miller, S.~Y.~Choi, B.~Eberle, M.~M.~Muhlleitner and P.~M.~Zerwas,
  Phys.\ Lett.\  B {\bf 505}, 149 (2001)
  [arXiv:hep-ph/0102023].

\bibitem{Schumacher:2001ax}
  M.~Schumacher,
  Note LC-PHSM-2001-003 (2001).

\bibitem{Djouadi:1995gt}
  A.~Djouadi, M.~Spira and P.~M.~Zerwas,
  Z.\ Phys.\  C {\bf 70}, 427 (1996)
  [arXiv:hep-ph/9511344].

\bibitem{Hildreth:1993dx}
  M.~D.~Hildreth, T.~L.~Barklow and D.~L.~Burke,
  Phys.\ Rev.\  D {\bf 49}, 3441 (1994).

\bibitem{Carena:2001bg}
  M.~Carena, H.~E.~Haber, H.~E.~Logan and S.~Mrenna,
  Phys.\ Rev.\  D {\bf 65}, 055005 (2002)
  [Erratum-ibid.\  D {\bf 65}, 099902 (2002)]
  [arXiv:hep-ph/0106116].

\bibitem{Desch:2004cu}
  K.~Desch, E.~Gross, S.~Heinemeyer, G.~Weiglein and L.~Zivkovic,
  JHEP {\bf 0409}, 062 (2004)
  [arXiv:hep-ph/0406322].

\bibitem{Battaglia:2004js}
  M.~Battaglia, D.~Dominici, J.~F.~Gunion and J.~D.~Wells,
  arXiv:hep-ph/0402062.

\bibitem{Battaglia:1999re}
  M.~Battaglia,
  arXiv:hep-ph/9910271.

\bibitem{Aubert:2004aw}
  B.~Aubert {\it et al.}  [BABAR Collaboration],
  \emph{Phys.\ Rev.\ Lett.\ } {\bf 93}, 011803 (2004)
  [arXiv:hep-ex/0404017].

\bibitem{Bauer:2002sh}
  C.~W.~Bauer, Z.~Ligeti, M.~Luke and A.~V.~Manohar,
  \emph{Phys.\ Rev.} {\bf D67}, 054012 (2003)
  [arXiv:hep-ph/0210027].

\bibitem{Battaglia:2002tm}
  M.~Battaglia {\it et al.},
  \emph{Phys.\ Lett.} {\bf B556}, 41 (2003)
  [arXiv:hep-ph/0210319].

\bibitem{Kuhl:2004ri}
  T.~Kuhl,
  prepared for the {\it International Conference on Linear Colliders (LCWS 04)}, 
  Paris, France, 19-24 April 2004.

\bibitem{Battaglia:2002gq}
  M.~Battaglia and A.~De Roeck,
  arXiv:hep-ph/0211207.

\bibitem{Battaglia:2002av}
  M.~Battaglia,
  arXiv:hep-ph/0211461.

\bibitem{Barklow:2003hz}
  T.~L.~Barklow,
  arXiv:hep-ph/0312268.

\bibitem{Djouadi:1999gv}
  A.~Djouadi, W.~Kilian, M.~Muhlleitner and P.~M.~Zerwas,
  \emph{Eur.\ Phys.\ J.} {\bf C10} (1999) 27
  [arXiv:hep-ph/9903229].

\bibitem{Kanemura:2004mg}
  S.~Kanemura, Y.~Okada, E.~Senaha and C.~P.~Yuan,
  Phys.\ Rev.\  D {\bf 70}, 115002 (2004)
  [arXiv:hep-ph/0408364].

\bibitem{Gutierrez-Rodriguez:2006qk}
  A.~Gutierrez-Rodriguez, M.~A.~Hernandez-Ruiz and O.~A.~Sampayo,
  arXiv:hep-ph/0601238.

\bibitem{Castanier:2001sf}
  C.~Castanier, P.~Gay, P.~Lutz and J.~Orloff,
  arXiv:hep-ex/0101028.

\bibitem{Battaglia:2001nn}
  M.~Battaglia, E.~Boos and W.~M.~Yao,
in {\it Proc. of the APS/DPF/DPB Summer Study on the Future of Particle Physics (Snowmass 2001) } ed. N.~Graf, E3016, [arXiv:hep-ph/0111276].

\bibitem{Baur:2002qd}
  U.~Baur, T.~Plehn and D.~L.~Rainwater,
  \emph{Phys.\ Rev.} {\bf D67}, 033003 (2003)
  [arXiv:hep-ph/0211224].

\bibitem{Baur:2003gp}
  U.~Baur, T.~Plehn and D.~L.~Rainwater,
  \emph{Phys.\ Rev.} {\bf D69}, 053004 (2004)
  [arXiv:hep-ph/0310056].

\bibitem{Barklow:2004th}
  T.~L.~Barklow,
  arXiv:hep-ph/0411221.

\bibitem{Datta:2005zs}
  A.~Datta, K.~Kong and K.~T.~Matchev,
  Phys.\ Rev.\ D {\bf 72}, 096006 (2005)
  [Erratum-ibid.\ D {\bf 72}, 119901 (2005)]
  [arXiv:hep-ph/0509246].

\bibitem{Smillie:2005ar}
  J.~M.~Smillie and B.~R.~Webber,
  JHEP {\bf 0510}, 069 (2005)
  [arXiv:hep-ph/0507170].

\bibitem{Battaglia:2005zf}
  M.~Battaglia, A.~Datta, A.~De Roeck, K.~Kong and K.~T.~Matchev,
  JHEP {\bf 0507}, 033 (2005)
  [arXiv:hep-ph/0502041].

\bibitem{wmap}
D. N. Spergel et al. [WMAP Collaboration], Astrophys. J. Suppl. {\bf 148}, 
175 (2003) [arXiv:astro-ph/0302209]

\bibitem{planck}
J. R. Bond, G. Efstathiou and M. Tegmark, Mon. Not. Roy. Astron. Soc. 
{\bf 291}, L33 (1997) [arXiv:astro-ph/9702100]

\bibitem{deBoer:2005tm}
  W.~de Boer, C.~Sander, V.~Zhukov, A.~V.~Gladyshev and D.~I.~Kazakov,
  Astron.\ Astrophys.\  {\bf 444}, 51 (2005)
  [arXiv:astro-ph/0508617].

\bibitem{Finkbeiner:2004us}
  D.~P.~Finkbeiner,
  arXiv:astro-ph/0409027.

\bibitem{Akerib:2005kh}
  D.~S.~Akerib {\it et al.}  [CDMS Collaboration],
  Phys.\ Rev.\ Lett.\  {\bf 96}, 011302 (2006)
  [arXiv:astro-ph/0509259].

\bibitem{Scherrer:1985zt}
  R.~J.~Scherrer and M.~S.~Turner,
  Phys.\ Rev.\ D {\bf 33}, 1585 (1986)
  [Erratum-ibid.\ D {\bf 34}, 3263 (1986)].

\bibitem{Battaglia:2003ab}
  M.~Battaglia, A.~De Roeck, J.~R.~Ellis, F.~Gianotti, K.~A.~Olive and L.~Pape,
  Eur.\ Phys.\ J.\ C {\bf 33}, 273 (2004)
  [arXiv:hep-ph/0306219].

\bibitem{Kong:2005hn}
  K.~Kong and K.~T.~Matchev,
  JHEP {\bf 0601}, 038 (2006)
  [arXiv:hep-ph/0509119].

\bibitem{Weiglein:2004hn}
  G.~Weiglein {\it et al.}  [LHC/LC Study Group],
  arXiv:hep-ph/0410364.

\bibitem{Gray:2005ci}
  R.~Gray {\it et al.},
  arXiv:hep-ex/0507008.

\bibitem{Khotilovich:2005gb}
  V.~Khotilovich, R.~Arnowitt, B.~Dutta and T.~Kamon,
  Phys.\ Lett.\ B {\bf 618}, 182 (2005)
  [arXiv:hep-ph/0503165].

\bibitem{Battaglia:2004gk}
  M.~Battaglia,
  arXiv:hep-ph/0410123.

\bibitem{ISAJET}
F.~E.~Paige, S.~D.~Protopescu, H.~Baer and X.~Tata,
arXiv:hep-ph/0312045.

\bibitem{Gondolo:2004sc}
P.~Gondolo, J.~Edsjo, P.~Ullio, L.~Bergstrom, M.~Schelke and E.~A.~Baltz,
JCAP {\bf 0407}, 008 (2004)
[arXiv:astro-ph/0406204].

\bibitem{Belanger:2006is}
  G.~Belanger, F.~Boudjema, A.~Pukhov and A.~Semenov,
  arXiv:hep-ph/0607059.

\bibitem{Baltz:2006fm}
  E.~A.~Baltz, M.~Battaglia, M.~E.~Peskin and T.~Wizansky,
  Phys.\ Rev.\ D {\bf 74}, 103521 (2006)
  [arXiv:hep-ph/0602187].

\bibitem{Feng:1993sd}
  J.~L.~Feng and D.~E.~Finnell, Phys.\ Rev.\ D {\bf 49}, 2369 (1994)
  [arXiv:hep-ph/9310211].

\bibitem{Moortgat-Pick:2005cw}
  G.~A.~Moortgat-Pick {\it et al.},
  arXiv:hep-ph/0507011, based on work of U.~Nauenberg {\it et al.}.

\bibitem{Blair:2001cz}
  G.~A.~Blair,
in {\it Proc. of the APS/DPF/DPB Summer Study on the Future of Particle Physics (Snowmass 2001) } ed. N.~Graf, E3019.

\bibitem{Martyn:2000}
H.~U.~Martyn and G.~A.~Blair, Note LC-TH-2000-023.

\bibitem{Bambade:2004tq}
  P.~Bambade, M.~Berggren, F.~Richard and Z.~Zhang,
  arXiv:hep-ph/0406010.

\bibitem{Chen:1989ss}
  P.~Chen and V.~I.~Telnov,
  Phys.\ Rev.\ Lett.\  {\bf 63}, 1796 (1989).

\bibitem{Tauchi:1993tm}
  T.~Tauchi, K.~Yokoya and P.~Chen,
  Part.\ Accel.\  {\bf 41}, 29 (1993).

\bibitem{Baer:2003ru}
  H.~Baer, A.~Belyaev, T.~Krupovnickas and X.~Tata,
  JHEP {\bf 0402}, 007 (2004)
  [arXiv:hep-ph/0311351].

\bibitem{Balazs:2004bu}
  C.~Balazs, M.~Carena and C.~E.~M.~Wagner,
  Phys.\ Rev.\  D {\bf 70}, 015007 (2004)
  [arXiv:hep-ph/0403224].

\bibitem{Feng:2005nz}
  J.~L.~Feng,
in Proc. of the {\it 2005 Int. Linear Collider Workshop (LCWS 2005)}, 
Stanford, California, 18-22 Mar 2005, pp 0013 and  [arXiv:hep-ph/0509309].

\bibitem{zp1}
M.~Battaglia {\it et al.}, in {\it Physics and Experiments with
Future Linear $e^+e^-$ Colliders}, (A.~Para and H.E.~Fisk editors), AIP Conference 
Proceedings, New York, 2001, 607 [arXix:hep-ph/0101114].

\bibitem{dominici}
D.~Dominici, arXiv:hep-ph/0110084.

\bibitem{Hewett:1993st}
  J.~L.~Hewett,
  arXiv:hep-ph/9308321.

\bibitem{Rizzo:2006nw}
  T.~G.~Rizzo,
  arXiv:hep-ph/0610104.

\bibitem{Ackerstaff:1997ke}
  K.~Ackerstaff {\it et al.}  [OPAL Collaboration],
  Z.\ Phys.\  C {\bf 75}, 385 (1997).

\bibitem{Abe:2004hx}
  K.~Abe {\it et al.}  [SLD Collaboration],
  Phys.\ Rev.\ Lett.\  {\bf 94}, 091801 (2005)
  [arXiv:hep-ex/0410042].

\bibitem{Hillert:2005rp}
  S.~Hillert  [LCFI Collaboration],
{\it In the Proceedings of 2005 International Linear Collider Workshop (LCWS 2005), Stanford, California, 18-22 Mar 2005, pp 0313}.

\bibitem{Chen:1993db}
  P.~Chen, T.~L.~Barklow and M.~E.~Peskin,
  Phys.\ Rev.\  D {\bf 49}, 3209 (1994)
  [arXiv:hep-ph/9305247].

\bibitem{Riemann:1997py}
  S.~Riemann,
  arXiv:hep-ph/9710564.

\bibitem{Battaglia:2001fr}
  M.~Battaglia, S.~De Curtis, D.~Dominici and S.~Riemann,
in {\it Proc. of the APS/DPF/DPB Summer Study on the Future of Particle Physics (Snowmass 2001) } ed. N.~Graf, E3020,  [arXiv:hep-ph/0112270].

\bibitem{Melles:2001ye}
  M.~Melles,
  Phys.\ Rept.\  {\bf 375}, 219 (2003)
  [arXiv:hep-ph/0104232].

\bibitem{Eichten:1983hw}
  E.~Eichten, K.~D.~Lane and M.~E.~Peskin,
  Phys.\ Rev.\ Lett.\  {\bf 50}, 811 (1983).

\bibitem{Ciafaloni:1999ub}
  P.~Ciafaloni and D.~Comelli,
  Phys.\ Lett.\ B {\bf 476}, 49 (2000)
  [arXiv:hep-ph/9910278].

\bibitem{Battaglia:2002ey}
  M.~Battaglia {\it et al.},
in {\it Proc. of the APS/DPF/DPB Summer Study on the Future of Particle Physics (Snowmass 2001) } ed. N.~Graf, E3006,  [arXiv:hep-ph/0201177].

\bibitem{concepts}
{\begin{verbatim}http://physics.uoregon.edu/~lc/wwstudy/concepts/ \end{verbatim}}

\bibitem{Behnke:2005re}
  T.~Behnke,
{\it In the Proceedings of 2005 International Linear Collider Workshop (LCWS 2005), Stanford, California, 18-22 Mar 2005, pp 0006}.

\bibitem{Abe:2006bz}
  K.~Abe {\it et al.}  [GLD Concept Study Group],
  arXiv:physics/0607154.

\bibitem{Battaglia:2006bv}
  M.~Battaglia, T.~Barklow, M.~Peskin, Y.~Okada, S.~Yamashita and P.~Zerwas,
{\it In the Proceedings of 2005 International Linear Collider Workshop (LCWS 2005), Stanford, California, 18-22 Mar 2005, pp 1602}
  [arXiv:hep-ex/0603010].

\bibitem{Abe:1999ky}
T.~Abe  [SLD Collaboration],
Nucl.\ Instrum.\ Meth.\ A {\bf 447} (2000) 90
[arXiv:hep-ex/9909048].

\bibitem{Turchetta:2001dy}
  R.~Turchetta {\it et al.},
  Nucl.\ Instrum.\ Meth.\ A {\bf 458} (2001) 677.

\bibitem{Marczewski:2005vy}
  J.~Marczewski {\it et al.},
  Nucl.\ Instrum.\ Meth.\ A {\bf 549} (2005) 112.

\bibitem{Richter:2003dn}
  R.~H.~Richter {\it et al.},
  Nucl.\ Instrum.\ Meth.\ A {\bf 511} (2003) 250.

\bibitem{Battaglia:2003kn}
  M.~Battaglia,
  Nucl.\ Instrum.\ Meth.\ A {\bf 530}, 33 (2004)
  [arXiv:physics/0312039].

\bibitem{pdg}
  W.-M. Yao {\it et al}, J.\ Phys.\ G {\bf 33}, 1 (2006)

\bibitem{Giomataris:1995fq}
  Y.~Giomataris, P.~Rebourgeard, J.~P.~Robert and G.~Charpak,
  Nucl.\ Instrum.\ Meth.\ A {\bf 376}, 29 (1996).

\bibitem{Sauli:1997qp}
  F.~Sauli,
  Nucl.\ Instrum.\ Meth.\ A {\bf 386}, 531 (1997).

\bibitem{Kappler:2004cg}
  S.~Kappler {\it et al.},
  IEEE Trans.\ Nucl.\ Sci.\  {\bf 51}, 1039 (2004).

\bibitem{Colas:2004ks}
  P.~Colas {\it et al.},
  Nucl.\ Instrum.\ Meth.\ A {\bf 535}, 506 (2004).

\bibitem{Kroseberg:2005ue}
  J.~Kroseberg {\it et al.},
  arXiv:physics/0511039.

\bibitem{Behnke:2001qq}
  T.~Behnke, S.~Bertolucci, R.~D.~Heuer and R.~Settles,
  \emph{TESLA Technical design report.  Pt. 4: A detector for TESLA}
DESY-01-011 (2001).

\bibitem{Brient:2002gh}
  J.~C.~Brient and H.~Videau,
in {\it Proc. of the APS/DPF/DPB Summer Study on the Future of Particle Physics (Snowmass 2001) } ed. N.~Graf, E3047,  [arXiv:hep-ex/0202004].

\bibitem{Buskulic:1994wz}
  D.~Buskulic {\it et al.}  [ALEPH Collaboration],
  Nucl.\ Instrum.\ Meth.\  A {\bf 360}, 481 (1995).

\bibitem{Videau:2000es}
  H.~Videau,
{\it Prepared for 5th International Linear Collider Workshop (LCWS 2000), Fermilab, Batavia, Illinois, 24-28 Oct 2000}

\bibitem{Strom:2005id}
  D.~Strom {\it et al.},
  IEEE Trans.\ Nucl.\ Sci.\  {\bf 52}, 868 (2005).

\bibitem{Mavromanolakis:2005yh}
  G.~Mavromanolakis,
{\it In the Proceedings of 2005 International Linear Collider Workshop (LCWS 2005), Stanford, California, 18-22 Mar 2005, pp 0906}
  [arXiv:physics/0510181].

\bibitem{Strom:2005xt}
  D.~Strom {\it et al.},
{\it In the Proceedings of 2005 International Linear Collider Workshop (LCWS 2005), Stanford, California, 18-22 Mar 2005, pp 0908}.

\end{thebibliography}
\end{document}